\documentclass[pra,twocolumn,showpacs,floatfix,a4paper,superscriptaddress]{revtex4}
\usepackage{graphicx}
\usepackage{amsmath}
\usepackage{amssymb}
\usepackage{amsfonts}
\usepackage{color}
\usepackage{subfigure}
\newcommand{\rmi}{\mathrm{i}}
\newcommand{\rme}{\mathrm{e}}
\newcommand{\rmd}{\mathrm{d}}
\newcommand{\rmm}{\mathrm{m}}
\newcommand{\rmc}{\mathrm{c}}
\newcommand{\rmin}{\mathrm{in}}
\newcommand{\mce}{\mathcal{E}}
\newcommand{\rmef}{\mathrm{eff}}

\begin{document}
\title{Effect of phase noise on the generation of stationary entanglement in cavity optomechanics}

\author{M. Abdi}
\affiliation{School of Science and Technology, Physics Division, University of Camerino, Camerino (MC), Italy}
\affiliation{Department of Physics, Sharif University of Technology, Tehran, Iran}
\author{Sh. Barzanjeh}
\affiliation{School of Science and Technology, Physics Division, University of Camerino, Camerino (MC), Italy}
\affiliation{Department of Physics, Faculty of Science, University of Isfahan, Hezar Jerib, 81746-73441, Isfahan, Iran }
\author{P. Tombesi}
\affiliation{School of Science and Technology, Physics Division, University of Camerino, Camerino (MC), Italy}
\affiliation{INFN, Sezione di Perugia, Italy}
\author{D. Vitali}
\affiliation{School of Science and Technology, Physics Division, University of Camerino, Camerino (MC), Italy}
\affiliation{INFN, Sezione di Perugia, Italy}

\begin{abstract}
We study the effect of laser phase noise on the generation of stationary entanglement between an intracavity optical mode and a mechanical resonator in a generic cavity optomechanical system. We show that one can realize robust stationary optomechanical entanglement even in the presence of non-negligible laser phase noise. We also show that the explicit form of the laser phase noise spectrum is relevant, and discuss its effect on both optomechanical entanglement and ground state cooling of the mechanical resonator.
\end{abstract}


\maketitle

%
%
\section{Introduction}

The study of optomechanics of micro- and nano-cavities has recently sparkled the interest of a broad scientific community due to its different applications, ranging from sensing of masses, forces and displacements at the ultimate quantum limits \cite{Schwab2005,Kippenberg2007}, to the realization of quantum interfaces for quantum information networks \cite{Mancini2003,Pirandola2004,Hammerer2009,Rabl2010}, up to tests of the validity of quantum mechanics at macroscopic level \cite{Marshall2003,Romero-Isart2011}.
The possibility to detect genuine quantum behavior in cavities characterized by an appreciable radiation pressure interaction between a light mode and a mechanical resonator was first pointed out by Braginski and coworkers \cite{Braginsky1995} in the context of the interferometric detection of gravitational waves. In the last years however many different schemes have been proposed for the detection of quantum mechanical effects in such systems, such as continuous variable (CV) entanglement between cavity modes and/or mechanical modes, squeezed states of the light or the mechanical modes, and ground state cooling of the mechanical modes (see Ref.~\cite{Genes2009} for a review). These schemes involve cavities and resonators at the micro- or nano-level rather than the macroscopic scale of gravitational wave detectors, and profit from the tremendous progress in micro- and nano-fabrication techniques which has provided novel opportunities to engineer optomechanical devices. Some examples are toroidal optical microresonators \cite{Kippenberg2007}, Fabry-Perot cavities with a movable micromirror \cite{Gigan2006,Arcizet2006}, a semitransparent membrane within a standard Fabry-Perot cavity \cite{Thompson2008,Jayich2008,Wilson2009,Sankey2010}, suspended silicon photonic waveguides~\cite{Li2008,Li2009,Li2009prl}, SiN nanowires evanescently coupled to a microtoroidal resonator \cite{Anetsberger2009}, adjacent photonic crystal wires \cite{Eichenfield2009}, nanoelectromechanical systems formed by a microwave cavity capacitively coupled to a nanoresonator \cite{Teufel2009,Rocheleau2010,Teufel2011}, atomic ensembles interacting with the mode of an optical cavity containing it \cite{Brennecke2008,Murch2008,Schleier-Smith2011}.

Here we focus on the possibility to generate robust CV entanglement between an intracavity optical mode and a mechanical resonator mode in the stationary state of the system, which has been predicted in \cite{Vitali2007} (see also Refs.~\cite{Genes2008b,Galve2010,Ghobadi2011} where the problem has been revisited, and Ref.~\cite{Vitali2007Milb} for its extension to the microwave cavity case). This steady-state entanglement could be very useful for quantum communication applications because it is endless, and robust against thermal noise. The above analyses however have not taken into account a possible technical limitation, associated with the fact that both the amplitude and especially the phase of the laser driving the cavity are noisy quantities. Laser phase noise could be in fact very dangerous for optomechanical entanglement, which is just the existence of strong quantum correlations between the fluctuations of cavity field quadratures at an appropriate phase and the position and momentum fluctuations of the mechanical resonator.

The effect of laser phase noise on ground state cooling of the mechanical resonator has been already discussed in Refs.~\cite{Diosi2008,Yin2009,Rabl2009,Phelps2011}. Ref.~\cite{Diosi2008} showed that phase noise acts on the mechanical resonator as an additional heating noise proportional to the intracavity field amplitude, that may represent a serious obstacle for ground state cooling. Ref.~\cite{Rabl2009} showed that what is relevant both for cooling and also for coherent state transfer between optical and mechanical modes is just the phase noise spectrum at the mechanical resonance frequency. If such a noise value is not too large, cooling is still possible as, in fact, has been confirmed experimentally in Refs.~\cite{Groblacher2009,Riviere2010} which approached, and Ref.~\cite{Teufel2011a} just reached, the ground state limit. Therefore the explicit form of the laser phase noise spectrum is relevant, and simply assuming a white phase noise tends to overestimates its effect because such a noise is strongly colored and decays significantly at MHz frequencies.

Here we extend the analysis to stationary optomechanical entanglement, in order to establish to what extent laser phase noise may affect its realization. Our analysis is based on a quantum Langevin equation treatment and generalizes previous approaches in various aspects. In particular it can deal with various examples of phase noise spectra, even though we shall focus on a bandpass filter form of the laser noise spectrum. This is in fact the typical case in current experiments because phase noise decays to zero at large frequencies, and it is negligible at low frequencies due to laser-cavity locking.
We show that laser phase noise has an appreciable effects on the achievable entanglement, but nonetheless significant stationary entanglement is still achievable employing currently available stabilized lasers. We also derive approximate analytical expression for the stationary optomechanical entanglement illustrating how laser phase noise affects its experimental realization. We shall also briefly reconsider the problem of mechanical ground state cooling.

The paper is organized as follows. In Sec.~II we present the model Hamiltonian and we show how one has to modify the standard quantum Langevin treatment in order to include in a non-perturbative way the effects of laser noise. In Sec.~III we study the linearized dynamics of quantum fluctuations around the appropriate classical steady states. In Sec.~IV we show the results for stationary optomechanical entanglement and cooling to the mechanical ground state. Sec.~V is for concluding remarks.

%
\section{Model}
%
%
We consider a generic cavity optomechanical system in which a mechanical resonator with frequency $\omega_{\rmm}$ is subject to a force proportional to the photon number of an optical cavity mode with frequency $\omega_{\rmc}$, which is driven by an intense, but noisy laser.
The corresponding Hamiltonian can be written as \cite{Kippenberg2007,Genes2009,Law1995,Giovannetti2001}
\begin{align}
H&=\hbar\omega_{\rmc}a^{\dagger}a+\frac{1}{2}\hbar\omega_{\rmm}(p^{2}+q^{2})
-\hbar G_{0}a^{\dagger}a q \nonumber \\
&+\rmi\hbar \mce(t)(a^{\dagger}\rme^{-\rmi[\omega_{0}t+\phi(t)]}-ae^{\rmi[\omega_{0}t+\phi(t)]}).
\label{hamiltonian}
\end{align}
The first term describes the energy of the cavity mode, with annihilation operator $a$ ($[a,a^{\dag}]=1$), while the second term gives the energy of the mechanical resonator, described by dimensionless position and momentum operators $q$ and $p$, satisfying the commutation relation $[q,p]=\rmi$.
The third term is the optomechanical interaction, with single photon optomechanical coupling strength
\begin{equation}\label{eq:coupling}
    G_0=-\left(\frac{\rmd \omega_{\rmc}}{\rmd x}\right)\sqrt{\frac{\hbar}{m \omega_{\rmm}}},
\end{equation}
where $(\rmd \omega_{\rmc}/\rmd x)$ is the change in cavity frequency per displacement and $m$ is the effective mass of the mechanical mode \cite{Rokhsari2006}.

The last term describes the cavity driving by a laser which is generally assumed to possess both phase and intensity fluctuations. The parameter $\omega_0$ denotes the laser average frequency and $\phi(t)$ is the zero-mean fluctuating phase, while $\mce(t)=\mce_0+\varepsilon(t)$ is related to the laser amplitude and $\varepsilon(t)$ describes the real, zero-mean amplitude fluctuations of the laser. The statistical properties of $\phi(t)$ and $\varepsilon(t)$ will be specified later on, while the mean amplitude is given by $\mce_0=\sqrt{2 \kappa P/\hbar \omega_0}$, where $P$ is the input laser power and $\kappa$ is the cavity loss rate through its input port.

The Hamiltonian of Eq.~(\ref{hamiltonian}) describes a wide variety of cavity optomechanical systems, with different cavity geometries and mechanical elements. In systems as Fabry-Perot cavities with a movable micro-mirror~\cite{Metzger2004,Gigan2006,Arcizet2006,Kleckner2006,Mow-Lowry2008}, or with a semitransparent membrane inside \cite{Thompson2008,Jayich2008,Wilson2009,Sankey2010}, or in radially vibrating toroidal microcavities~\cite{Schliesser2010}, optomechanical coupling is provided by radiation pressure. In other optomechanical devices coupling is instead provided by the transverse gradient force, such as in suspended silicon photonic waveguides \cite{Li2008,Li2009,Li2009prl}, SiN nanowire evanescently coupled to a microtoroidal resonator \cite{Anetsberger2009}, and in "zipper" cavities formed by two adjacent photonic crystal wires \cite{Eichenfield2009}.
The same Hamiltonian applies also to nanoelectromechanical systems formed by a microwave cavity capacitively coupled to a nanoresonator, such as in \cite{Teufel2009,Rocheleau2010,Teufel2011}, and in such a case the noisy laser describes the phase and intensity fluctuations of the microwave driving source. Finally Eq.~(\ref{hamiltonian}) also applies to systems where a mechanical collective degree of freedom of an atomic ensemble interacts with an optical cavity containing it \cite{Brennecke2008,Murch2008,Schleier-Smith2011}. In all these devices one always has many cavity and mechanical modes, but one can adopt the Hamiltonian of Eq.~(\ref{hamiltonian}) whenever one can restrict to single cavity and mechanical modes. This is justified when the cavity free spectral range is much larger than the mechanical frequency $\omega_{\rmm}$ (i.e., not too large cavities): in such a case the input laser drives only one cavity mode and scattering of photons from the driven mode into other cavity modes is negligible \cite{Law1995}. One can restrict to a single mechanical mode
when the detection bandwidth is chosen so that it includes only a single, isolated, mechanical resonance and mode-mode coupling is negligible~\cite{Genes2008c}.

\begin{figure}[t]
\centering
\includegraphics[width=3.4in]{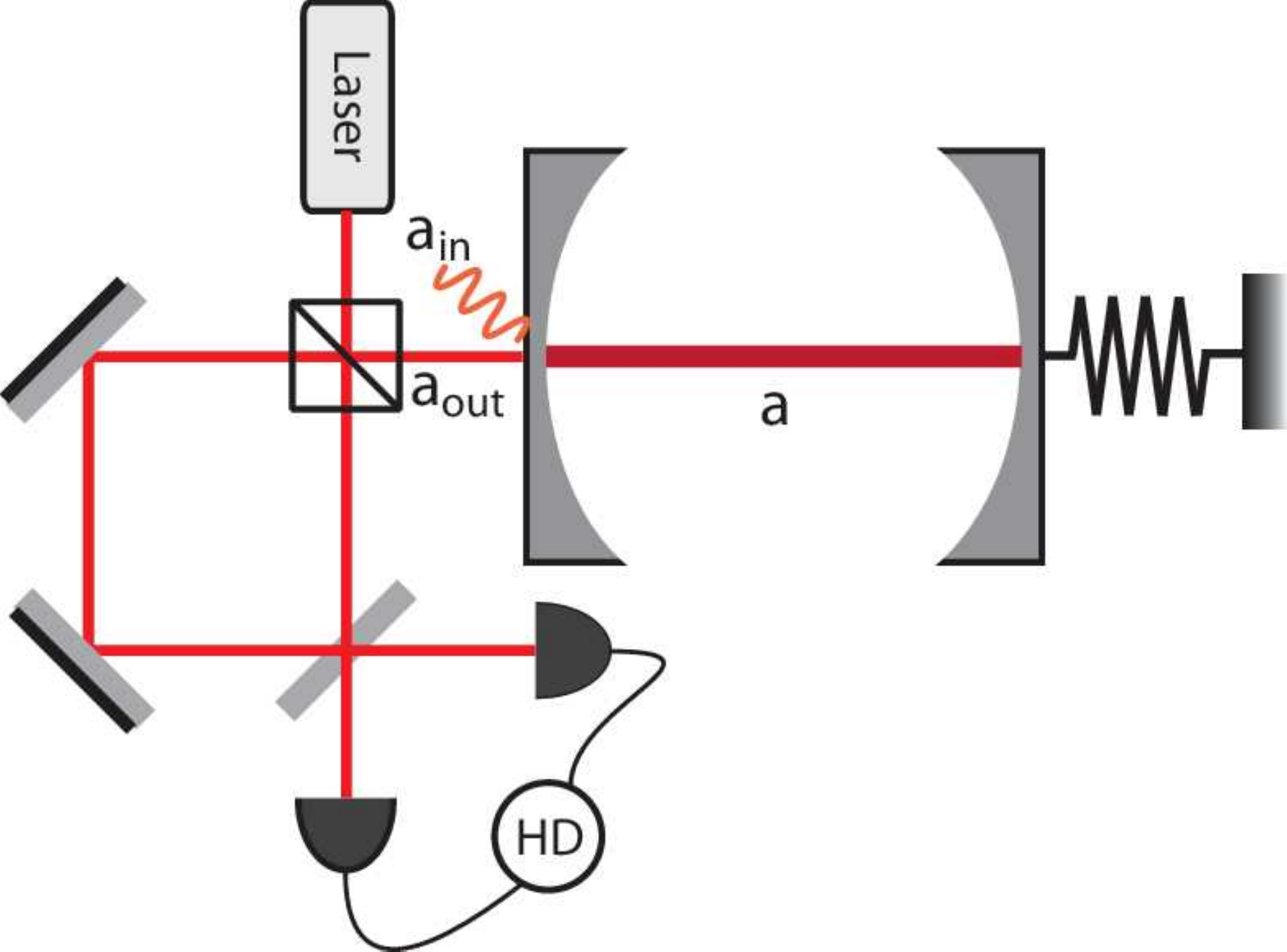}
\caption{
Scheme of the cavity optomechanical system under study monitored by an homodyne detection apparatus.}
\label{fig:homodyne}
\end{figure}
\subsection*{Quantum Langevin equations}

For a full description of the system dynamics it is necessary to include the fluctuation-dissipation processes affecting both the optical and the mechanical mode. They can be taken into account in a fully consistent way \cite{Giovannetti2001}, and one gets
\begin{subequations}
\label{nonlinear0}
\begin{eqnarray}
\dot{q}&=&\omega_{\rmm} p, \\
\dot{p}&=&-\omega_{\rmm} q - \gamma_{\rmm} p + G_0 a^{\dag}a + \xi, \\
\dot{a}&=&-[\kappa+\rmi\omega_{\rmc}]a +i G_0 a q \nonumber \\
&&+\mce(t)\rme^{-\rmi[\omega_{0}t+\phi(t)]}+\sqrt{2\kappa} a_{\rmin},
\end{eqnarray}
\end{subequations}
where $a_{\rmin}$ is the vacuum input noise, whose nonzero correlations are given by~\cite{Gardiner2000}
\begin{subequations}
\label{input}
\begin{eqnarray}
\langle a_{\rmin}(t)a^{\dag}_{\rmin}(t')\rangle &=& \left[N(\omega_{\rmc})+1\right] \delta (t-t'),\\
\langle a^{\dag}_{\rmin}(t)a_{\rmin}(t')\rangle &=& N(\omega_{\rmc})\delta (t-t'),
\end{eqnarray}
\end{subequations}
with $N(\omega_{\rmc})=\left(\exp\{\hbar \omega_{\rmc}/k_{\mathrm{B}}T\}-1\right)^{-1}$ the equilibrium mean thermal photon number ($k_{\mathrm{B}}$ is the Boltzmann constant and $T$ is the temperature of the reservoir).
At optical frequencies $\hbar \omega_{\rmc}/k_{\mathrm{B}}T \gg 1$ and therefore $N(\omega_{\rmc})\approx 0$, so that only the correlation function of Eq.~(\ref{input}a) is relevant. We have assumed for simplicity the ideal case of a single-ended cavity, so that the total cavity decay rate coincides with the loss rate through the input port $\kappa$. The mechanical mode is affected by a viscous force with damping rate $\gamma_{\rmm}$ and by a Brownian stochastic force with zero mean value $\xi(t)$, obeying the correlation function \cite{Landau1958,Gardiner2000}
\begin{equation}
\left \langle \xi(t) \xi(t')\right \rangle = \frac{\gamma_{\rmm}}{\omega_{\rmm}}%
\int \frac{\rmd \omega}{2\pi} \rme^{-\rmi \omega(t-t')} \omega \left[\coth\left(\frac{\hbar \omega}{2k_{\mathrm{B}}T}\right)+1\right].
\label{brownian}
\end{equation}
The Brownian noise $\xi(t)$ is a Gaussian quantum stochastic process and its non-Markovian nature (neither its correlation function nor its commutator are proportional to a Dirac delta function) guarantees that the QLE of Eqs.~(\ref{nonlinear}) preserve the correct commutation relations between operators during the time evolution \cite{Giovannetti2001}.

We are interested in the entanglement properties of the stationary state of the system, which are determined by the quantum correlations between the mechanical and optical field quadratures. These quadratures can be detected by homodyning the cavity output and an additional weak field probing the mechanical element (see for example Ref.~\cite{Vitali2007}). The local oscillator for the homodyne detector is provided just by the noisy driving laser (see Fig.~\ref{fig:homodyne}) and this means that all detected quantities are referred to the frame rotating at the \emph{fluctuating} instantaneous frequency $\omega_0+\dot{\phi}(t)$. Passing to this randomly rotating frame, the cavity field operator transforms according to $a(t) \to a(t)\exp\left\{-\rmi\omega_{0}(t-t_0)-\rmi\int_{t_0}^t dt'\dot{\phi}(t')\right\}$, where $t_0 \to -\infty$ is the time instant at which we fix the phase reference for the cavity field by taking $\mce(t)$ real; as a consequence Eqs.~(\ref{nonlinear0}) become
\begin{subequations}
\label{nonlinear}
\begin{eqnarray}
\dot{q}&=&\omega_{\rmm} p, \\
\dot{p}&=&-\omega_{\rmm} q - \gamma_{\rmm} p + G_0 a^{\dag}a + \xi, \\
\dot{a}&=&-\kappa a -\rmi(\Delta_0-\dot{\phi}-G_0 q)a   +\mce(t)+\sqrt{2\kappa} \tilde{a}_{\rmin}(t),
\end{eqnarray}
\end{subequations}
where $\Delta_0=\omega_{\rmc}-\omega_0$ is the detuning of the cavity mode from the average laser frequency, and $ \tilde{a}_{\rmin}(t)=a_{\rmin}(t)\rme^{\rmi\omega_{0}(t-t_0)+\rmi\int_{t_0}^t dt'\dot{\phi}(t')}$ is still vacuum input noise, possessing the same correlation functions of $a_{\rmin}(t)$ [see Eqs.~(\ref{input})]. Therefore in the frame rotating at the fluctuating frequency, laser amplitude noise acts as additive noise on the cavity modes, while laser frequency noise is a multiplicative noise, affecting the cavity field in the same manner of the fluctuations of the resonator position $q$.

%
%
\section{Linearized dynamics of the fluctuations}

Achieving stationary optomechanical entanglement means establishing strong quantum correlations between the steady-state fluctuations of the position and momentum of the resonator, and the intracavity field quadrature fluctuations. As shown in Refs.~\cite{Vitali2007,Genes2008b}, this is attained when the effective coupling between these fluctuations is strong, which is realized when the intracavity field is very intense,
i.e., for high-finesse cavities and enough driving power. Therefore we focus onto the dynamics of the fluctuations around the classical steady state.

When the system is stable, it reaches a steady state which, in the absence of laser noise, is characterized by the cavity mode in a coherent state, and the mechanical resonator at an equilibrium position shifted by a quantity proportional to the stationary intracavity photon number. One may expect that, due to laser phase noise, this classical steady state is modified: in particular one expects that the phase of the intracavity coherent state slowly becomes completely random. This is the classical steady state assumed in Refs.~\cite{Rabl2009,Phelps2011}, i.e., a cavity field in a coherent state with completely random phase but with time-independent photon number, so that the shift of the equilibrium position of the resonator is not changed. Our treatment adopts however the frame rotating at the fluctuating instantaneous laser frequency $\omega_0+\dot{\phi}(t)$, differently from Refs.~\cite{Rabl2009,Phelps2011} which adopt the frame rotating at $\omega_0$. In such a fluctuating frame the phase of the classical stationary coherent state is not random, and its amplitude $\alpha_{\mathrm{s}}$ is given by the standard expression valid in the absence of laser noise, $\alpha_{\mathrm{s}} = \mce_0/(\kappa+\rmi \Delta$). This latter equation is actually an implicit nonlinear equation for $\alpha_{\mathrm{s}}$ because
\begin{equation}\label{eq:delta}
\Delta = \Delta_0- G_0 q_{\mathrm{s}}=\Delta_0- \frac{G_0^2 |\alpha_{\mathrm{s}}|^2}{\omega_{\rmm}},
\end{equation}
the effective cavity detuning, depends upon $|\alpha_{\mathrm{s}}|^2$. Assuming such a time-independent classical steady state is equivalent to assume that both phase and amplitude noise of the driving laser affect only the quantum fluctuations of the system and not its classical stationary state. In fact, inserting such a steady state solution into Eqs.~(\ref{nonlinear}), the exact (nonlinear) QLE for the fluctuations become
\begin{subequations}
\label{semilinear}
\begin{eqnarray}
\delta\dot{q}&=&\omega_{\rmm} \delta p,  \\
\delta\dot{p}&=&-\omega_{\rmm} \delta q - \gamma_{\rmm} \delta p + G_0 (\alpha_{\mathrm{s}} \delta a^{\dag}+\alpha_{\mathrm{s}}^{*} \delta a)+ \delta a^{\dag} \delta a + \xi, \\
\delta\dot{a}&=&-[\kappa+\rmi\Delta] \delta a+\rmi G_0 \alpha_{\mathrm{s}}\delta q+\rmi[G_0 \delta q +\dot{\phi}]\delta a \nonumber \\
&&+\rmi \dot{\phi}\alpha_{\mathrm{s}} +\varepsilon+\sqrt{2\kappa} \tilde{a}_{\rmin}.
\end{eqnarray}
\end{subequations}
As discussed above, robust optomechanical entanglement can be generated when the effective coupling between the fluctuations $G_0 \alpha_{\mathrm{s}}$ is large enough, which is best achieved when $|\alpha_{\mathrm{s}}| \gg 1$, i.e., we have a large number of intracavity photons. In such a case the system dynamics is well described by linearizing Eqs.~(\ref{semilinear}), i.e., by neglecting the term $\delta a^{\dag} \delta a$ in Eq.~(\ref{semilinear}b) and the terms $\rmi[G_0 \delta q +\dot{\phi}]\delta a$ in  Eq.~(\ref{semilinear}c), so that
\begin{subequations}
\label{linear}
\begin{eqnarray}
\delta\dot{q}&=&\omega_{\rmm} \delta p,  \\
\delta\dot{p}&=&-\omega_{\rmm} \delta q - \gamma_{\rmm} \delta p + G_0 (\alpha_{\mathrm{s}} \delta a^{\dag}+\alpha_{\mathrm{s}}^{*} \delta a)+ \xi,  \\
\delta\dot{a}&=&-[\kappa+\rmi\Delta] \delta a+\rmi G_0 \alpha_{\mathrm{s}}\delta q+\rmi \dot{\phi}\alpha_{\mathrm{s}} +\varepsilon+\sqrt{2\kappa} \tilde{a}_{\rmin}.
\end{eqnarray}
\end{subequations}
Notice that in this way we neglect also the multiplicative noise term $\rmi \dot{\phi} \delta a$ together with the usual nonlinear terms, but this is reasonable because when $|\alpha_{\mathrm{s}}| \gg 1$ such a term has a negligible effect compared to that of the $i \dot{\phi}\alpha_{\mathrm{s}}$ term. It is convenient to rewrite the linearized QLE in terms of the field quadrature fluctuations $\delta X_\Delta \equiv\left(\delta a \rme^{\rmi \theta_\Delta} +\delta a^{\dag}\rme^{-\rmi \theta_\Delta} \right)/\sqrt{2}$ and $\delta Y_\Delta \equiv\left(\delta a \rme^{\rmi \theta_\Delta} -\delta a^{\dag}\rme^{-\rmi \theta_\Delta} \right)/\rmi \sqrt{2}$ ,
\begin{subequations}
\label{linearquadra}
\begin{eqnarray}
\delta\dot{q}&=&\omega_{\rmm} \delta p,  \\
\delta\dot{p}&=&-\omega_{\rmm} \delta q - \gamma_{\rmm} \delta p + G \delta X_\Delta +\xi,\\
\delta\dot{X}_\Delta &=&-\kappa \delta X_\Delta +\Delta \delta Y_\Delta +\sqrt{2} \cos \theta_\Delta \varepsilon  +\sqrt{2\kappa} X_\Delta ^{\rmin},\\
\delta\dot{Y}_\Delta &=&-\kappa \delta Y_\Delta -\Delta \delta Y_\Delta +G \delta q \nonumber \\
&&+ \sqrt{2}|\alpha_{\mathrm{s}}|\dot{\phi}+\sqrt{2} \sin \theta_\Delta \varepsilon  +\sqrt{2\kappa} Y_\Delta ^{\rmin},
\end{eqnarray}
\end{subequations}
where $\theta_{\Delta} = \arctan [\Delta/\kappa]$, $G=G_0 \sqrt{2}|\alpha_{\mathrm{s}}|$ is the effective optomechanical coupling, and we have introduced the corresponding Hermitian input noise operators $X^{\rmin}_{\Delta}\equiv(\tilde{a}_{\rmin}\rme^{\rmi \theta_\Delta}+\tilde{a}^{\dag}_{\rmin}\rme^{-\rmi \theta_\Delta})/\sqrt{2}$ and $%
Y^{\rmin}_{\Delta}\equiv(\tilde{a}_{\rmin}\rme^{\rmi \theta_\Delta}-\tilde{a}^{\dag}_{\rmin}\rme^{-\rmi \theta_\Delta})/\rmi\sqrt{2}$.

Finally we have to specify the statistical properties of laser phase and amplitude noise. In currently available stabilized lasers, amplitude noise $\varepsilon(t)$ is negligible with respect to phase noise $\phi(t)$ and therefore we shall neglect it from now on, as assumed also in Refs.~\cite{Diosi2008,Yin2009,Rabl2009,Phelps2011}. Phase noise instead is typically non-negligible and it is responsible for the nonzero laser linewidth $\Gamma_l$. In fact, the laser spectrum is given by the Fourier transform of the stationary correlation function of the field, that is,
\begin{equation}\label{eq:lasspectrum}
   \mathcal{S}_{\rm L}(\omega)=\int \rmd\tau \rme^{\rmi \omega \tau} C(\tau) = \int \rmd\tau \rme^{\rmi \omega \tau}\langle \exp\{\rmi\phi(t+\tau)-\rmi\phi(t)\}\rangle ;
\end{equation}
$\phi(t)$ is well described by a zero-mean stationary Gaussian stochastic process, and therefore one can write
\begin{eqnarray}\label{eq:corrfunc}
    C(\tau)&=& \left \langle\exp\left\{\rmi\int_0^{\tau} {\rm d}s \dot{\phi}(s)\right\}\right\rangle \nonumber \\
    & = &\exp\left\{-\frac{1}{2}\int_0^{\tau}{\rm d}s \int_0^{\tau}{\rm d}s' \langle \dot{\phi}(s) \dot{\phi}(s')\rangle\right\}.
\end{eqnarray}
If one takes a delta correlated frequency noise $\langle \dot{\phi}(s) \dot{\phi}(s')\rangle = 2 \Gamma_l \delta (s-s')$, i.e., a flat frequency noise spectrum $\mathcal{S}_{\dot{\phi}}(\omega)=2\Gamma_l$, Eq.~(\ref{eq:corrfunc}) yields  $C(\tau)=\rme^{-\Gamma_l |\tau|}$, which corresponds to a Lorenzian laser spectrum with linewidth $\Gamma_l$. However, as already pointed out in Ref.~\cite{Rabl2009}, taking a flat $\mathcal{S}_{\dot{\phi}}(\omega)$ tends to overestimate the effect of laser phase noise; in practice the frequency noise spectrum has a bandpass filter form and therefore the laser spectrum is no more a perfect Lorenzian, but has faster decaying tails. A more realistic description is obtained by taking the following bandpass filter form of the frequency noise spectrum
\begin{equation}\label{eq:freqspectrum}
   \mathcal{S}_{\dot{\phi}}(\omega)=2\Gamma_l \frac{\Omega^4}{(\Omega^2-\omega^2)^2+\omega^2 \tilde{\gamma}^2},
\end{equation}
with $\Omega$ denoting the band center and $\tilde{\gamma}$ the bandwidth of the frequency noise spectrum, while the noise strength $\Gamma_l$ still describes the laser linewidth. A flat frequency noise spectrum is recovered in the limit $\Omega \to \infty$, $\tilde{\gamma} \to \infty$. It is straightforward to verify that the frequency noise spectrum of Eq.~(\ref{eq:freqspectrum}) is reproduced by assuming that the frequency noise variable $\psi\equiv \dot{\phi}$ satisfies the following pair of Langevin equations
\begin{subequations}
\label{phase}
\begin{eqnarray}
\dot{\psi}&=&\Omega \theta, \\
\dot{\theta}&=&-\Omega \psi-\tilde{\gamma}\theta+\Omega \sqrt{2\Gamma_l} \epsilon(t),
\end{eqnarray}
\end{subequations}
where $\epsilon(t)$ is a white noise with correlation function $\langle \epsilon(t)\epsilon(t') \rangle_{\mathrm{cl}} = \delta (t-t')$.
If we now attach these two latter equations to the linearized QLE of Eqs.~(\ref{linearquadra}), i.e., we treat the variables $\psi$ and $\theta$ as two additional dynamical variables of the system, we get that the system dynamics in the presence of laser phase noise is fully described by the following set of equations in matrix form
\begin{equation}
\dot{u}(t)=A u(t)+n(t),
\label{compact}
\end{equation}
where $u(t) =(\delta q(t), \delta p(t),\delta X_{\Delta}(t), \delta Y_{\Delta}(t), \psi(t), \theta(t))^{\mathsf{T}}$ is the vector of CV fluctuation operators, and $n(t) =(0, \xi(t),\sqrt{2\kappa}X_{\Delta}^{\rmin}(t), \sqrt{2\kappa}Y_{\Delta}^{\rmin}(t), 0, \Omega \sqrt{2\Gamma_l} \epsilon(t))^{\mathsf{T}}$ is the corresponding vector of noises.
Moreover, the drift matrix $A$ is the $6 \times 6$ matrix
\begin{equation}
A=\left(\begin{array}{cccccc}
    0 & \omega_{\rmm} & 0 & 0 & 0 & 0 \\
    -\omega_{\rmm} & -\gamma_{\rmm} & G & 0 & 0 & 0 \\
    0 & 0 & -\kappa & \Delta & 0 & 0 \\
    G & 0 & -\Delta & -\kappa & \sqrt{2}|\alpha_{\mathrm{s}}| & 0\\
    0 & 0 & 0 & 0 & 0 & \Omega \\
    0 & 0 & 0 & 0 & -\Omega & -\tilde{\gamma}
  \end{array}\right).
\label{drift}
\end{equation}

\subsection{Stationary quantum fluctuations}

We are interested in the stationary properties of the system: in particular we want to check to what extent laser phase noise hinders achieving a steady state with distinct quantum properties, in particular characterized by CV entanglement between the cavity mode and the mechanical element. The realization of such a stationary optomechanical entanglement is of particular interest for quantum information applications, because it would represent a very robust source of persistent entanglement.

The steady state associated with Eq.~(\ref{compact}) is reached when the system is stable, which occurs if and only if all the eigenvalues of $A$ have negative real part. These stability conditions can be obtained for example by using the Routh-Hurwitz criteria, and it is possible to verify that they are not modified by the presence of laser phase noise. Therefore they coincide with those discussed in Refs.~\cite{Vitali2007,Genes2008b,Genes2008}; in this paper we shall restrict to the situation with $\Delta > 0$, i.e., with a red-detuned laser, and in this parameter region the only non-trivial stability condition is $G^2 < \left(\Delta^2+\kappa^2\right)\omega_{\rmm}/\Delta$.

The steady state is a zero-mean Gaussian state due to the fact that the dynamics of the fluctuations is linearized and all noises are Gaussian;
as a consequence, it is fully characterized by the $6 \times 6$ stationary correlation matrix (CM) $V$, with matrix elements
\begin{equation}
 V_{ij}=\frac{\langle u_i(\infty)u_j(\infty)+ u_j(\infty)u_i(\infty)\rangle}{2}.
\label{cm1}
\end{equation}
The formal solution of Eq.~(\ref{compact}) yields \cite{Genes2008b}
\begin{equation}\label{v1}
V_{ij}=\int_0^{\infty} \rmd s \int_0^{\infty}\rmd s' M_{ik}(s) M_{jl}(s')D_{kl}(s-s'),
\end{equation}
where $M(t)=\exp(A t)$ and $D(s-s')$ is the diffusion matrix, the matrix of noise correlations, defined as $D_{kl}(s-s')=\langle n_k(s)n_l(s')+n_l(s')n_k(s) \rangle/2$.
The Brownian noise $\xi(t)$ is in general a non-Markovian Gaussian noise (see Eq.~(\ref{brownian})), but in the limit of large mechanical
quality factor $Q_{\rmm}=\omega_{\rmm}/\gamma_{\rmm} \gg 1$, becomes with a good approximation Markovian, with symmetrized correlation function
\begin{equation}
\label{eq:browncorre2}
\frac{\langle \xi(t)\xi(t^{\prime})+\xi(t^{\prime})\xi(t)\rangle}{2} \simeq \gamma_{\rmm}(2n+1) \delta(t-t^{\prime}),
\end{equation}
where $n=\left( \exp \{\hbar \omega _{\rmm}/k_{B}T\}-1\right) ^{-1}$ is the mean thermal phonon number at $T$.
Therefore, the diffusion matrix becomes $D(s-s')=D\delta(s-s')$, where $D=\mathrm{diag}[0,\gamma_{\rmm} (2n+1),\kappa,\kappa,0,2 \Gamma_l \Omega^2]$, so that Eq.~(\ref{v1}) simplifies to
\begin{equation}
V =\int_0^{\infty} \rmd s  M(s)D M(s)^{\mathsf{T}},
\label{cm3}
\end{equation}
which, when the stability conditions are satisfied ($M(\infty)=0$), is equivalent to the following Lyapunov equation for $V$
\begin{equation}
AV+VA^{\mathsf{T}}=-D.
\label{lyapunov}
\end{equation}
Eq.~(\ref{lyapunov}) is a linear equation for $V$ and can be straightforwardly solved, but its explicit expression is cumbersome and will not be reported here.

%
%
\section{Results}
From the solution of Eq.~(\ref{lyapunov}) for the stationary CM $V$, we can determine all the quantum properties of the stationary state of the cavity optomechanical system. We determine in particular the effect of laser phase noise on the possibility to achieve optomechanical entanglement and ground state cooling of the mechanical resonator.
\subsection{Entanglement}
The auxiliary variables $\psi$ and $\theta$ do not refer to the optomechanical system of interest and therefore we are concerned with
the reduced $4 \times 4$ CM extracted from Eq.~(\ref{cm3}) by neglecting the last two rows and columns.
This reduced correlation matrix can be expressed in the following form
\begin{equation}
V \equiv \left(\begin{array}{cc}
    V_{\mathrm{A}} & V_{\mathrm{C}} \\
    V_{\mathrm{C}}^{\mathsf{T}} & V_{\mathrm{B}}
  \end{array}\right),
\label{cm4}
\end{equation}
where $V_{\mathrm{A}}$, $V_{\mathrm{B}}$ and $V_{\mathrm{C}}$ are $2 \times 2$ matrices, with
$V_{\mathrm{A}}$ associated to the mechanical resonator, $V_{\mathrm{B}}$ to the cavity mode, and $V_{\mathrm{C}}$ describing the optomechanical correlations.
A convenient measure for CV entanglement is the logarithmic negativity \cite{Vidal2002,Adesso2004}, given by
\begin{equation}
E_{N}=\mathrm{max}(0, -\ln 2\eta^-),
\end{equation}
where $\eta^-$ is symplectic eigenvalue of the bipartite system and it is given by the equation
\begin{equation}
\eta^- \equiv \frac{1}{\sqrt{2}} \big[\Sigma(V)-\sqrt{\Sigma(V)^2 - 4 \det V} \big]^{1/2},
\end{equation}
with $\Sigma(V)=\det V_{\mathrm{A}}+\det V_{\mathrm{B}}-2 \det V_{\mathrm{C}}$.

We now study the behavior of $E_{N}$ when the phase noise parameters, i.e., the laser linewidth $\Gamma_l$ and the center of the phase noise spectrum $\Omega$, are varied. For simplicity we shall always consider a bandpass filter form of the phase noise spectrum, i.e., Eq.~(\ref{eq:freqspectrum}) with $\tilde{\gamma}=\Omega/2$. We consider typical parameter values, i.e., a mechanical resonator with $\omega_{\rmm}/2\pi=10$ MHz, quality factor $Q_{\rmm}=2 \times 10^6$, and at $T=0.4$ K. We also assume a single photon optomechanical coupling strength $G_0=1$ KHz, which is achieved for example in a Fabry-Perot cavity with length $L=1$ mm and with an oscillating micromirror of effective mass $m \simeq 10$ ng. However, comparable or even larger values of $G_0$ are currently achieved in other cavity optomechanics setups.
\begin{figure*}[t]
\centering
\subfigure{
\includegraphics[width=2.33in]{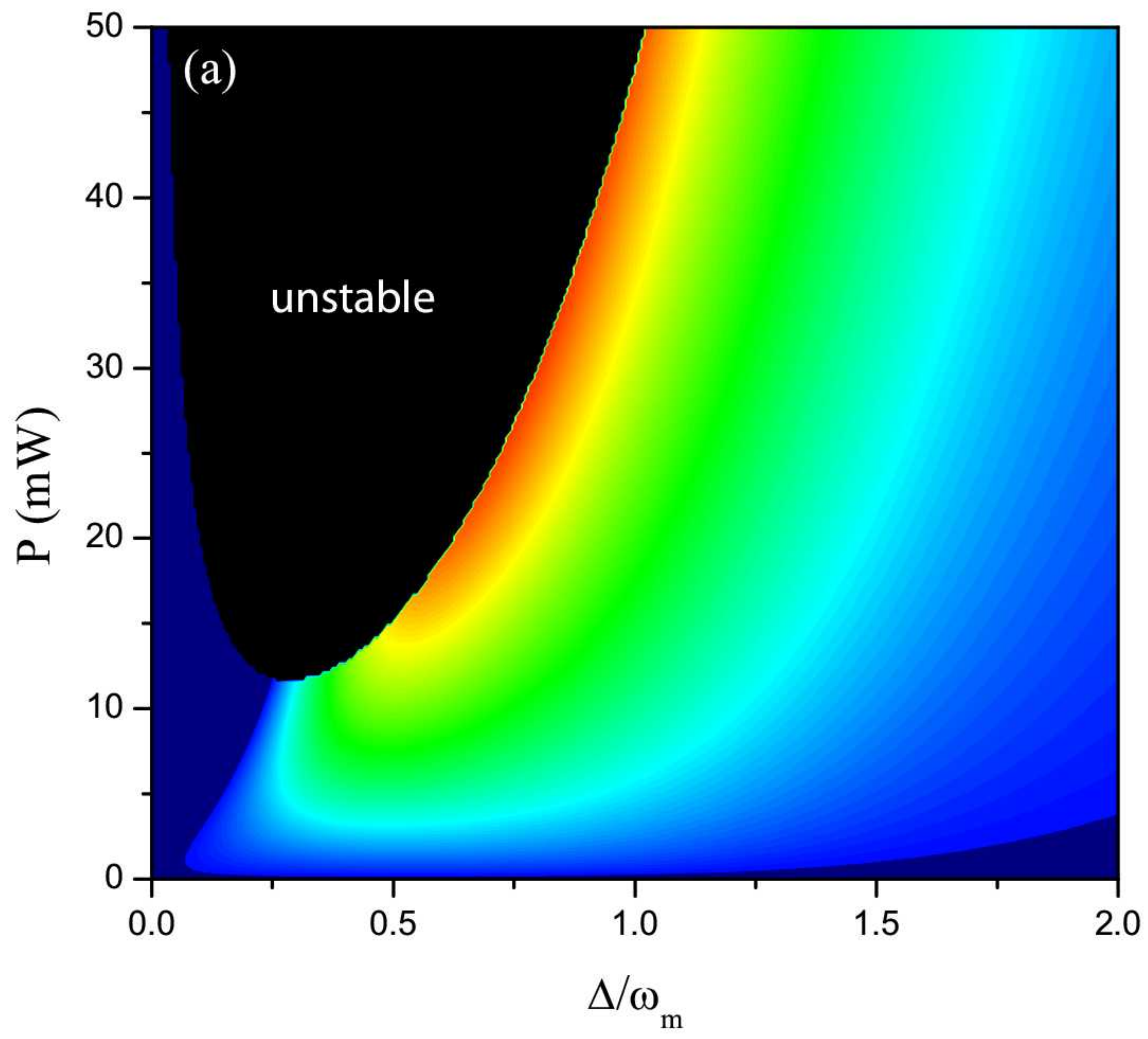}
\label{fig:entdetlin_a}
}
\subfigure{
\includegraphics[width=2.08in]{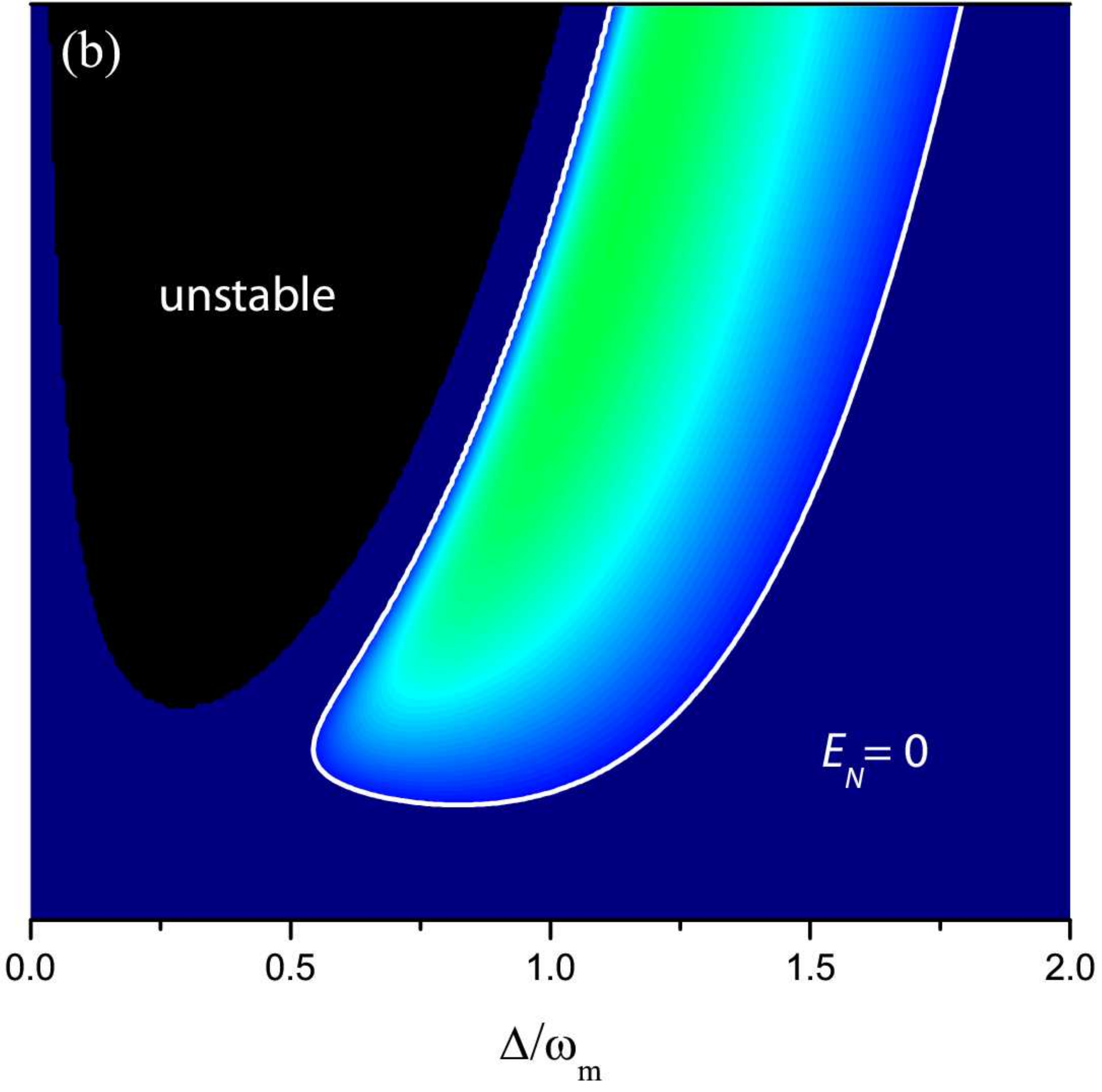}
\label{fig:entdetlin_b}
}
\subfigure{
\includegraphics[width=2.31in]{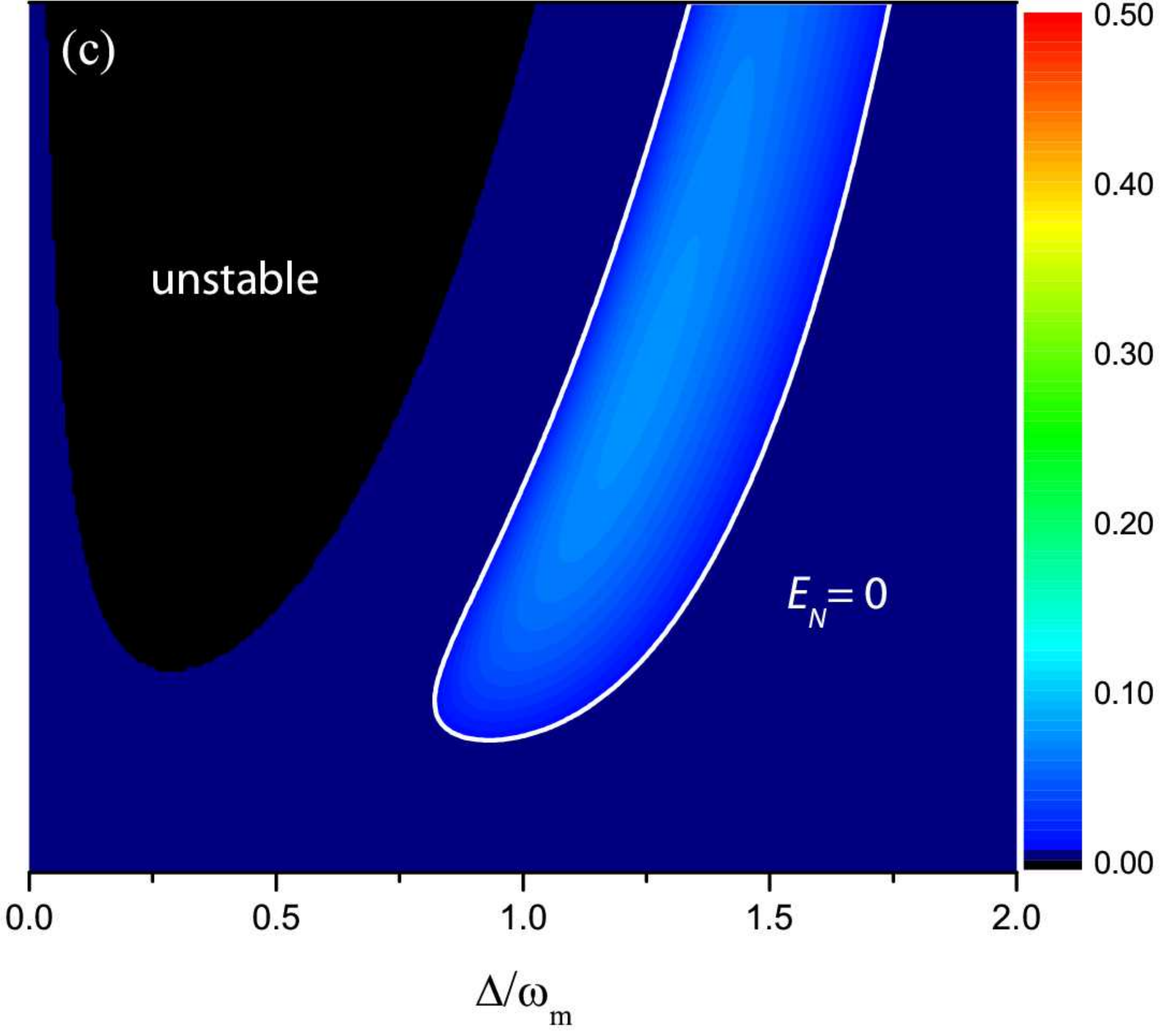}
\label{fig:entdetlin_c}
}
\caption{(Color online) Contour plot of $E_{N}$ versus the input power $P$ and normalized cavity detuning $\Delta/\omega_{\rmm}$ for $\Gamma_l=0$ (a) (no phase noise), $\Gamma_l/2\pi=0.1$ KHz (b), and $\Gamma_l/2\pi=1$ KHz (c). The cavity bandwidth has been fixed at $\kappa/\omega_{\rmm}=0.5$ and we have also fixed the center of the phase noise band $\Omega/2\pi=50$ KHz. See text for the other parameter values.}
\label{fig:entdetlin}
\end{figure*}
\begin{figure*}[t]
\centering
\subfigure{
\includegraphics[width=2.36in]{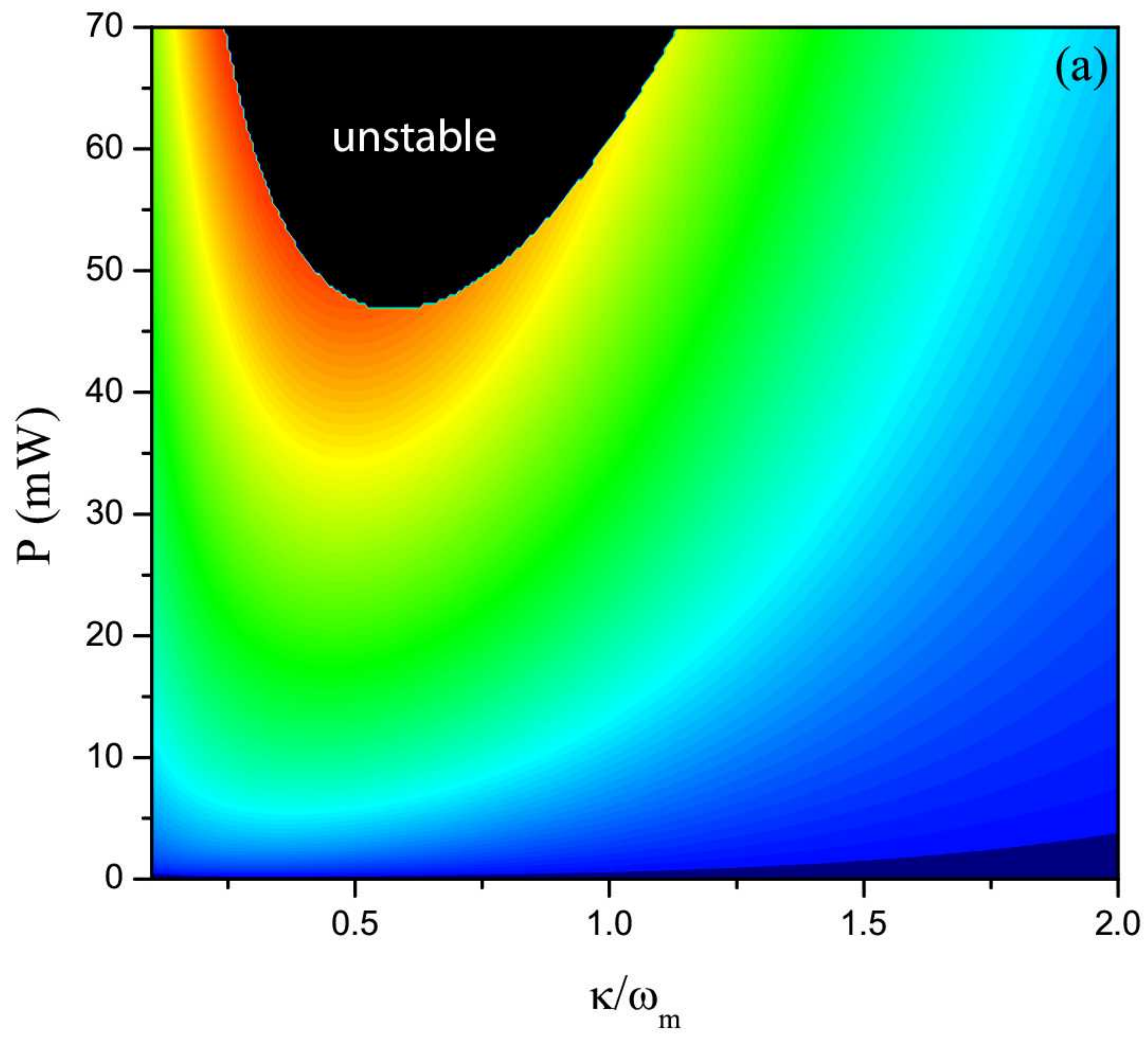}
\label{fig:entfinlin_a}
}
\subfigure{
\includegraphics[width=2.05in]{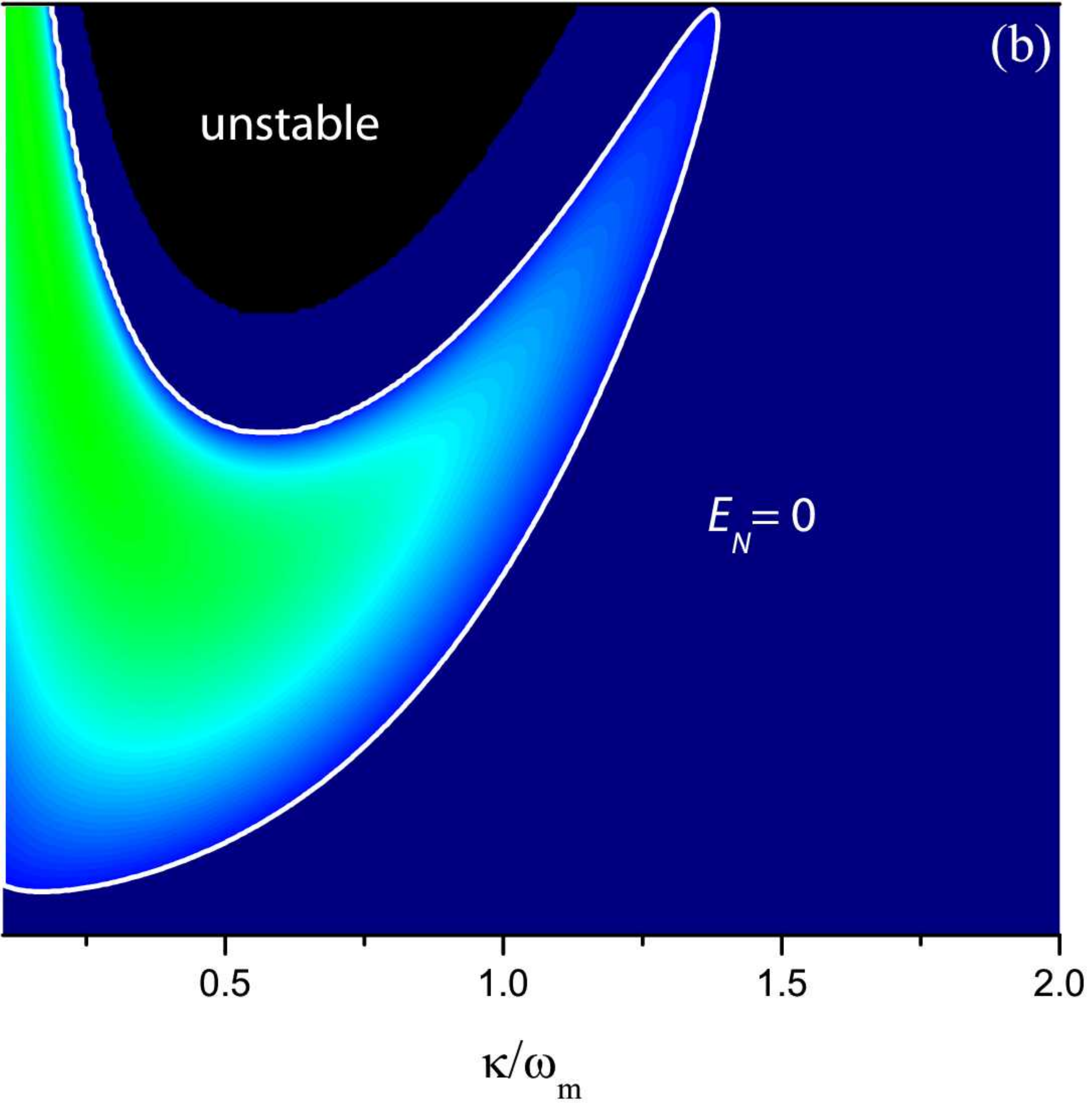}
\label{fig:entfinlin_b}
}
\subfigure{
\includegraphics[width=2.29in]{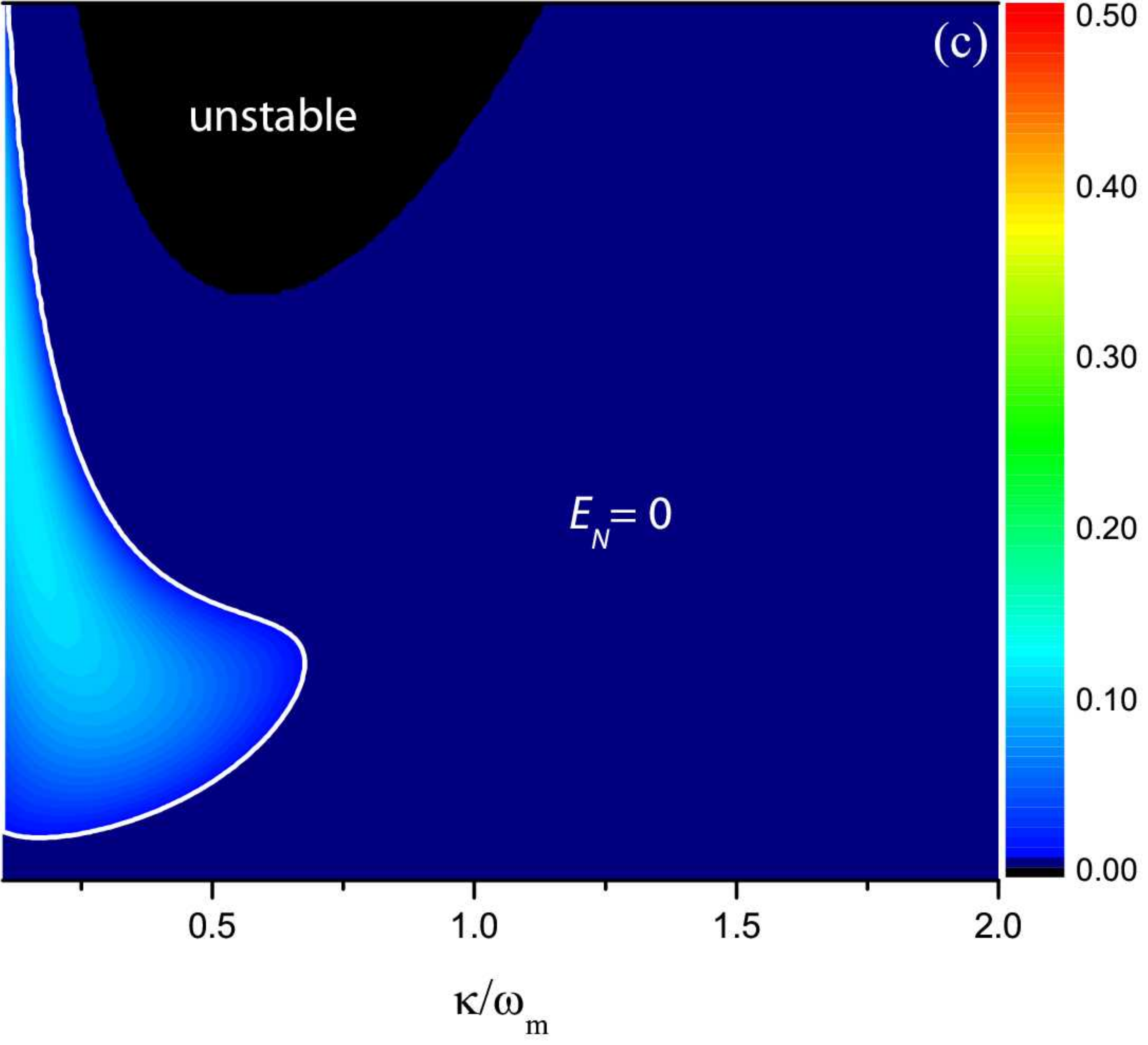}
\label{fig:entfinlin_c}
}
\caption{(Color online) Contour plot of $E_{N}$ versus the input power $P$ and normalized cavity decay rate $\kappa/\omega_{\rmm}$ for $\Gamma_l=0$ (a), $\Gamma_l/2\pi=0.1$ KHz (b), and $\Gamma_l/2\pi=1$ KHz (c). The cavity detuning has been fixed at $\Delta/\omega_{\rmm}=1$ and we have fixed $\Omega/2\pi=50$ KHz. See text for the other parameter values.}
\label{fig:entfinlin}
\end{figure*}

Typical values of the linewidth of stabilized lasers are around $\Gamma_l/2\pi \simeq 1$ KHz and therefore it is important to see if the stationary optomechanical entanglement predicted in Refs.~\cite{Vitali2007,Genes2008b} is robust against laser phase noise. In Fig.~\ref{fig:entdetlin} we show $E_{N}$ versus the input power $P$ and normalized cavity detuning $\Delta/\omega_{\rmm}$ for $\Gamma_l=0$ (a) (no phase noise), $\Gamma_l/2\pi=0.1$ KHz (b), and $\Gamma_l/2\pi=1$ KHz (c). The cavity bandwidth has been fixed at $\kappa/\omega_{\rmm}=0.5$ and we have also fixed the center of the phase noise spectrum $\Omega/2\pi=50$ KHz. We see that laser phase noise has an appreciable effect on the log-negativity: for increasing $\Gamma_l$ its maximum value decreases, and the parameter region where the steady state is entangled significantly narrows. Phase noise in particularly destructive close to the instability threshold, where $E_N$ is maximum when  $\Gamma_l=0$: as soon as $\Gamma_l \neq 0$, entanglement vanishes at the threshold for bistability and the maximum value of $E_N$ is achieved far from the threshold, still around $\Delta/\omega_{\rmm} \simeq 1$ and at intermediate values of the input power. In fact, in the presence of phase noise it is no more helpful to increase the input power because not only $G$ but also $|\alpha_{\rm s}|$ becomes larger, amplifying in this way the effect of phase noise [see Eqs.~(\ref{linear}c) and (\ref{linearquadra}d)].

Fig.~3 shows instead $E_{N}$ versus the input power $P$ and normalized cavity decay rate $\kappa/\omega_{\rmm}$ for $\Gamma_l=0$ (a), $\Gamma_l/2\pi=0.1$ KHz (b), and $\Gamma_l/2\pi=1$ KHz (c). The cavity detuning has been fixed at $\Delta/\omega_{\rmm}=1$ and we have again fixed $\Omega/2\pi=50$ KHz. The destructive effects of laser phase noise manifest again by decreasing the maximum achievable $E_N$ and narrowing the parameter region with a nonzero entanglement. Again $E_N$ vanishes at the bistability threshold, and for increasing phase noise entanglement is found only in the resolved sideband region, $\kappa/\omega_{\rmm} < 1$.

\begin{figure*}
\centering
\subfigure{
\includegraphics[width=2.33in]{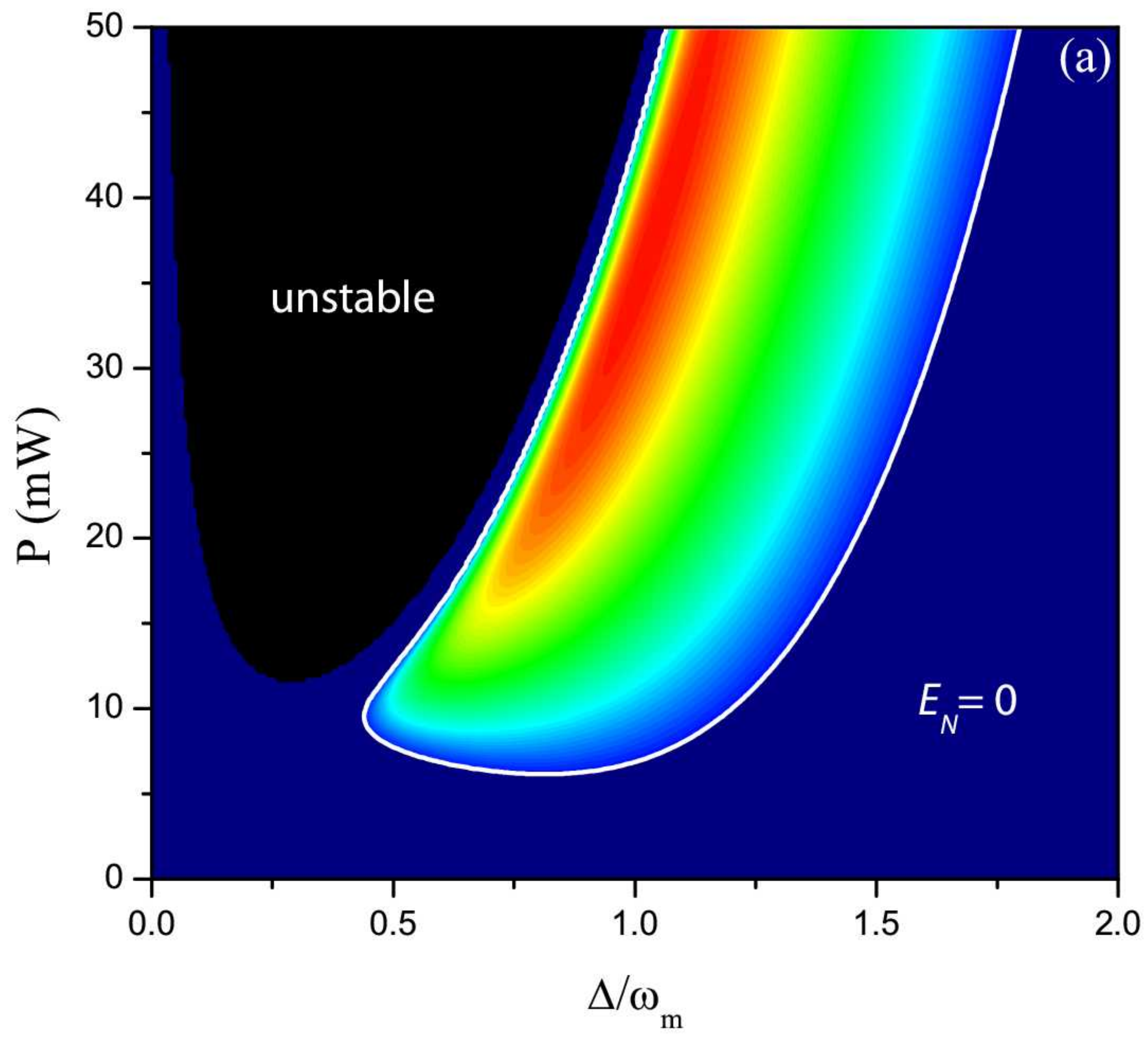}
\label{fig:entdetcut_a}
}
\subfigure{
\includegraphics[width=2.08in]{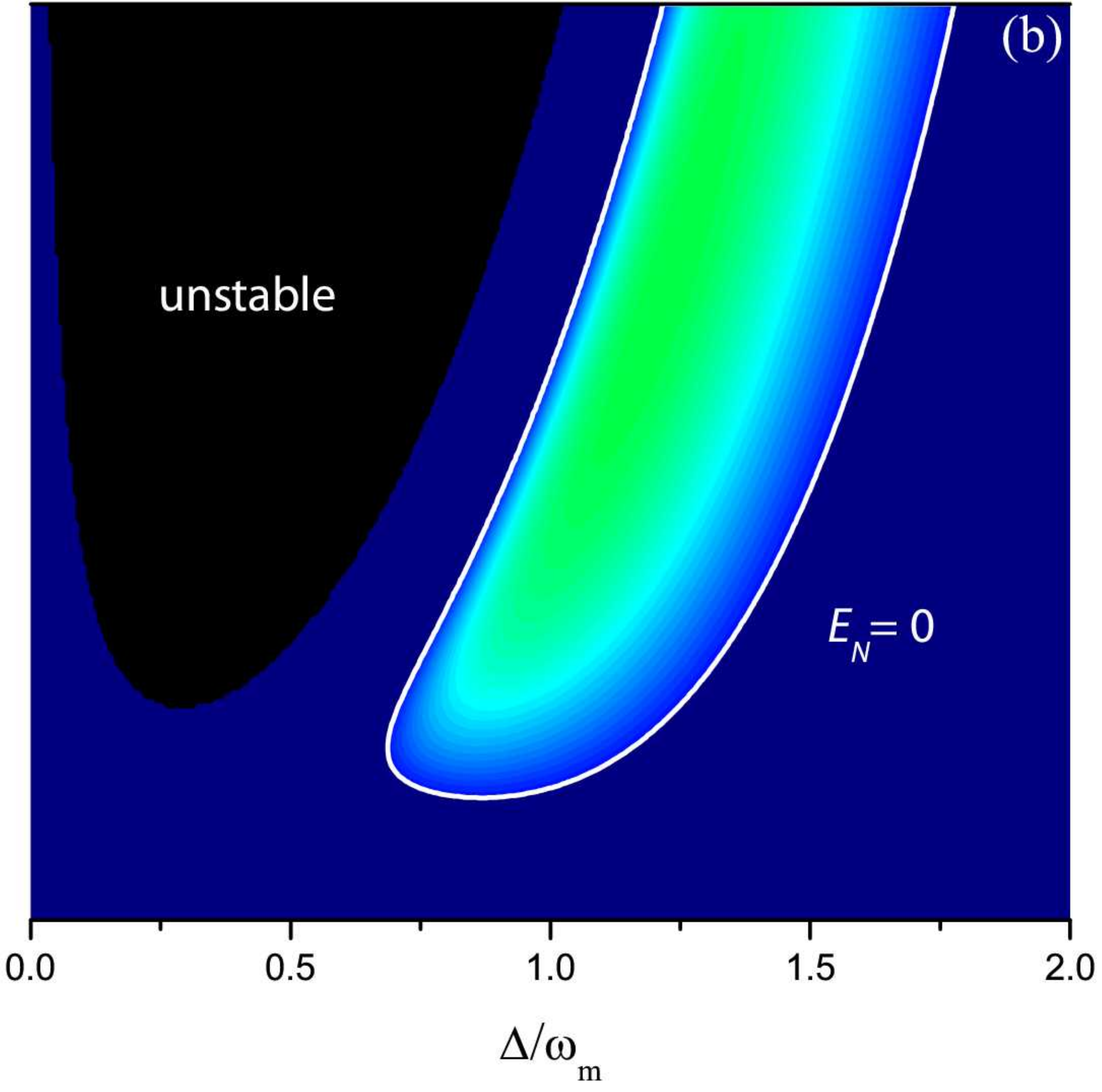}
\label{fig:entdetcut_b}
}
\subfigure{
\includegraphics[width=2.31in]{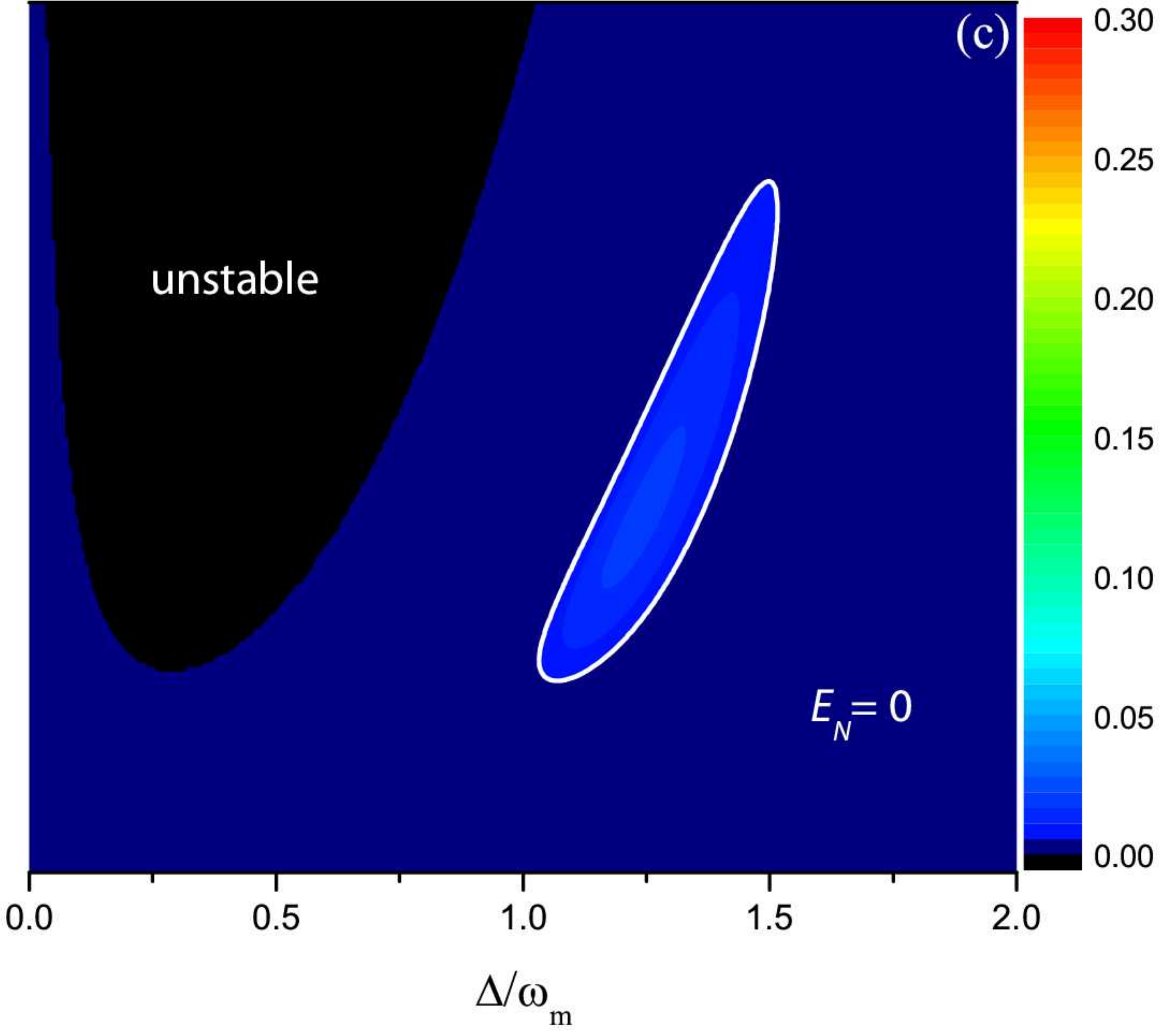}
\label{fig:entdetcut_c}
}
\caption{(Color online) Contour plot of $E_{N}$ versus the input power $P$ and normalized cavity detuning $\Delta/\omega_{\rmm}$ at fixed laser linewidth $\Gamma_l/2\pi=0.1$ KHz, and for different values of the center of the frequency noise band, $\Omega/2\pi =30$ KHz (a), $\Omega/2\pi =80$ KHz (b), and $\Omega/2\pi =140$ KHz (c) (the bandwidth parameter $\tilde{\gamma}$ is correspondingly adjusted so that $\tilde{\gamma}=\Omega/2$). The cavity decay rate has been fixed at $\kappa/\omega_{\rmm}=0.5$. See text for the other parameter values.}
\label{fig:entdetcut}
\end{figure*}
\begin{figure*}
\centering
\subfigure{
\includegraphics[width=2.36in]{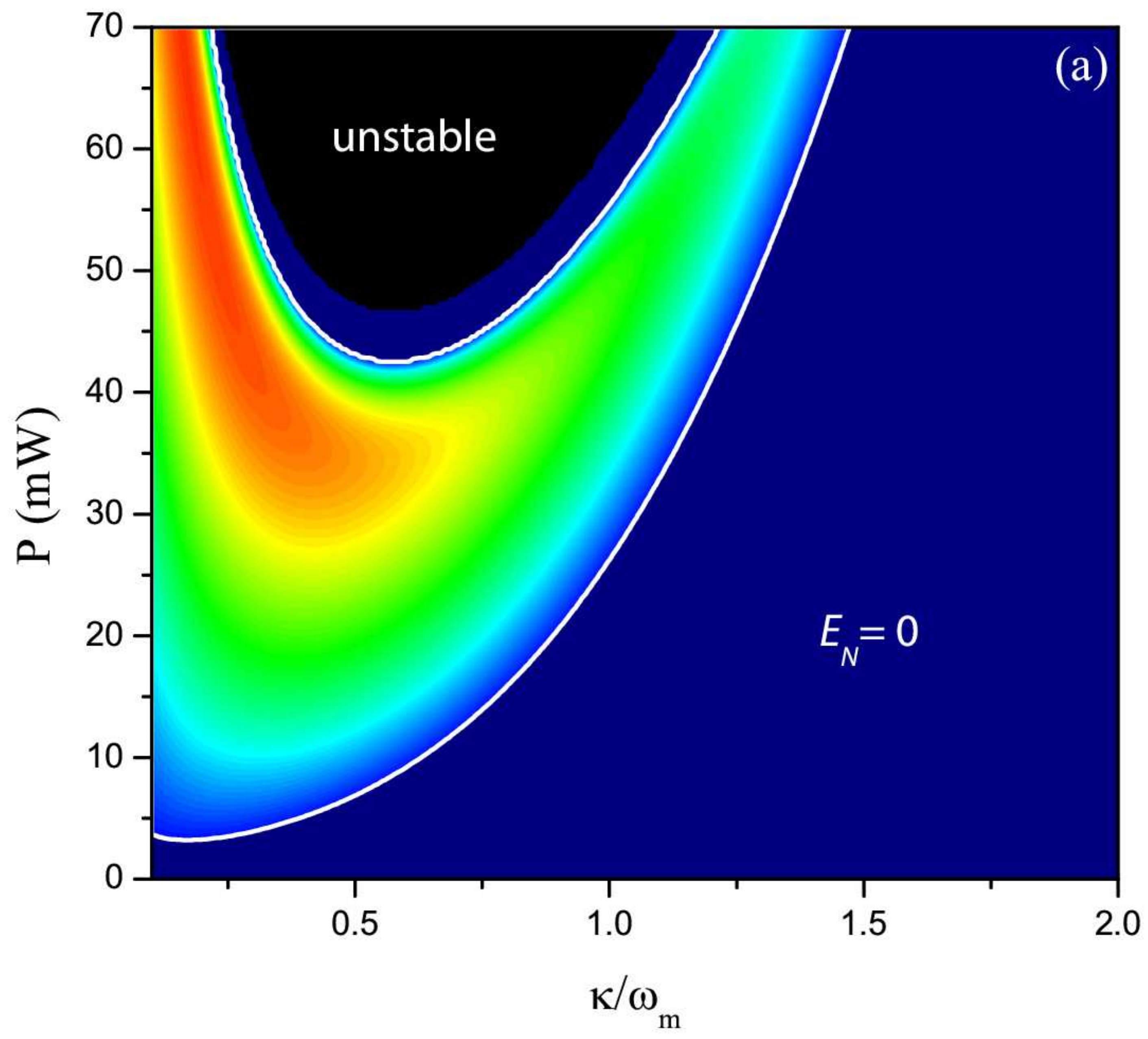}
\label{fig:entfincut_a}
}
\subfigure{
\includegraphics[width=2.05in]{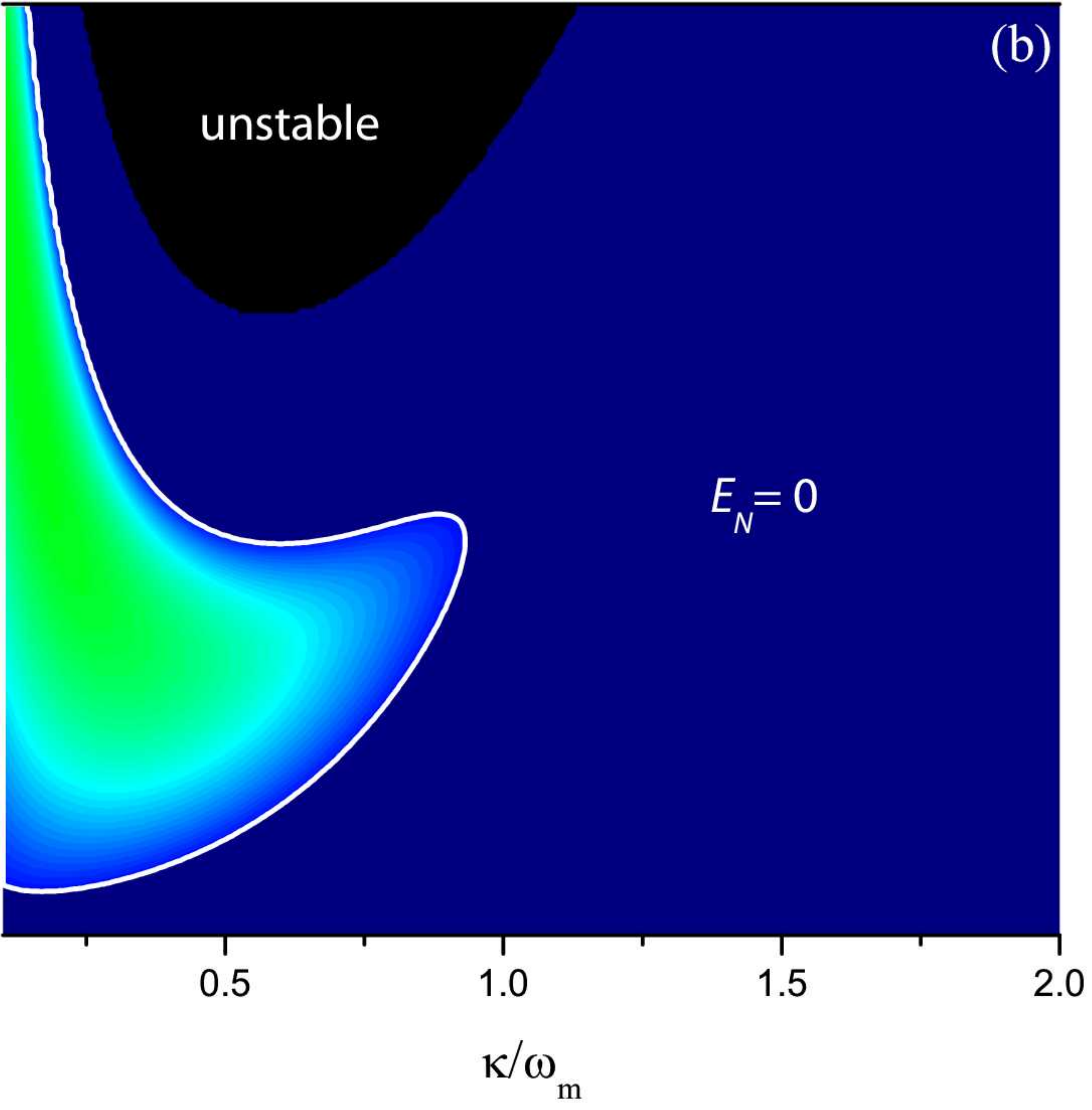}
\label{fig:entfincut_b}
}
\subfigure{
\includegraphics[width=2.29in]{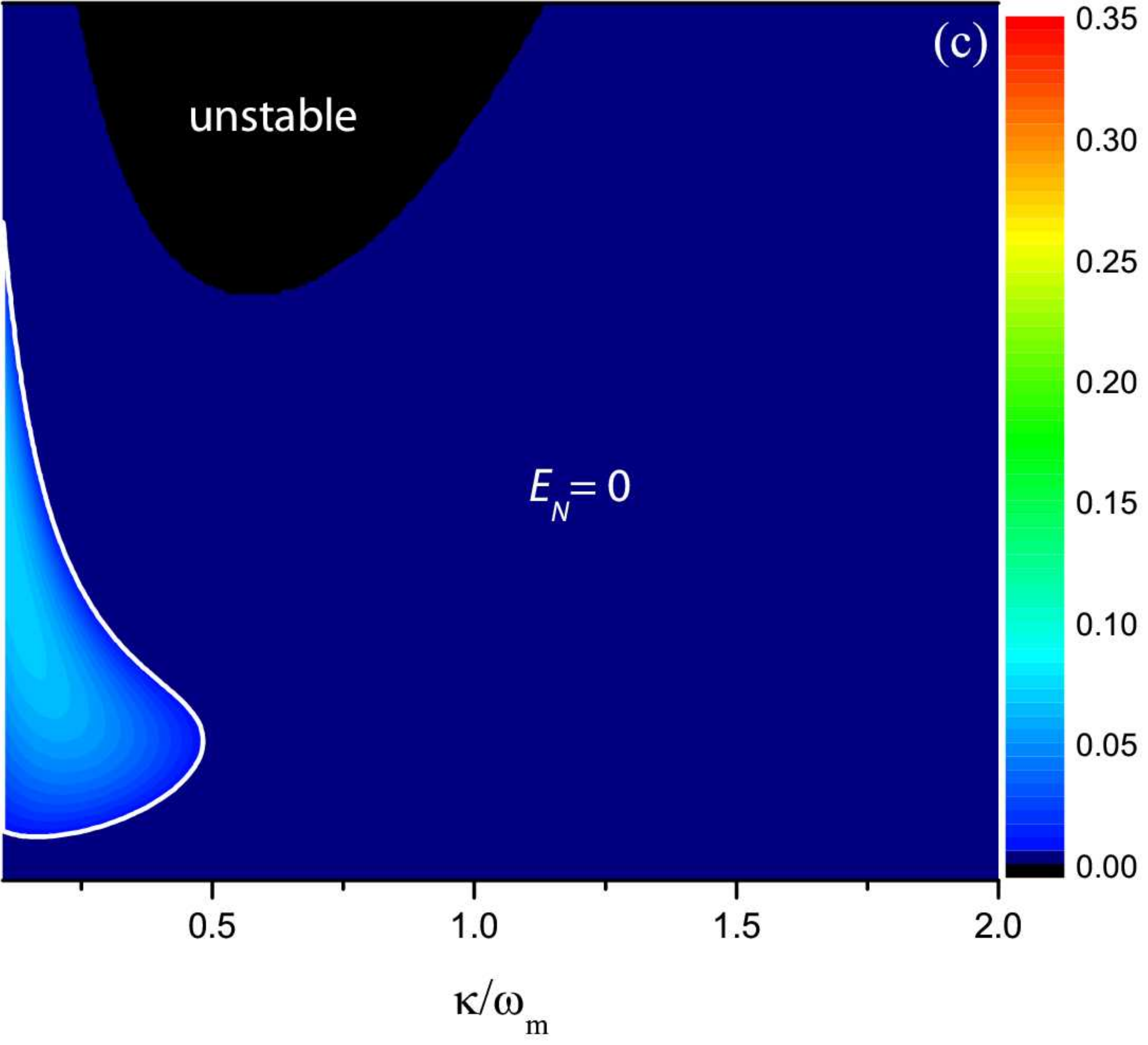}
\label{fig:entfincut_c}
}
\caption{Contour plot of $E_{N}$ versus the input power $P$ and normalized cavity decay rate $\kappa/\omega_{\rmm}$ at fixed laser linewidth $\Gamma_l/2\pi=0.1$ KHz, and for different values of the center of the frequency noise band, $\Omega/2\pi =30$ KHz (a), $\Omega/2\pi =80$ KHz (b), and $\Omega/2\pi =140$ KHz (c) (the bandwidth parameter $\tilde{\gamma}$ is correspondingly adjusted so that $\tilde{\gamma}=\Omega/2$). The detuning has been fixed at $\Delta/\omega_{\rmm}=1$. See text for the other parameter values.}
\label{fig:entfincut}
\end{figure*}

In Fig.~4 instead we study the dependence of entanglement upon the spectral properties of laser phase noise. It shows $E_{N}$ versus the input power $P$ and normalized cavity detuning $\Delta/\omega_{\rmm}$ at fixed laser linewidth $\Gamma_l/2\pi=0.1$ KHz, and for different values of the center of the frequency noise band, $\Omega/2\pi =30$ KHz (a), $\Omega/2\pi =80$ KHz (b), and $\Omega/2\pi =140$ KHz (c) (the bandwidth parameter $\tilde{\gamma}$ is correspondingly adjusted so that $\tilde{\gamma}=\Omega/2$). The cavity decay rate has been fixed at $\kappa/\omega_{\rmm}=0.5$. The figures clearly show that the spectral properties of frequency noise have a strong effect on stationary entanglement, which in fact progressively worsens for broader and broader frequency noise spectrum. In fact entanglement is still considerable when $\Omega/2\pi = 30$ KHz and $\tilde{\gamma}/2\pi=15$ KHz, but becomes extremely small when $\Omega/2\pi = 140$ KHz and $\tilde{\gamma}/2\pi=70$ KHz, even with a moderate phase noise strength, $\Gamma_l/2\pi=0.1$ KHz.

The relevance of the form of the noise spectrum is also evident in Fig.~5, where we show $E_{N}$ versus the input power $P$ and normalized cavity decay rate $\kappa/\omega_{\rmm}$ at fixed laser linewidth $\Gamma_l/2\pi=0.1$ KHz, again for $\Omega/2\pi =30$ KHz (a), $\Omega/2\pi =80$ KHz (b), and $\Omega/2\pi =140$ KHz (c). The cavity detuning has been fixed at $\Delta/\omega_{\rmm}=1$  and again the bandwidth parameter $\tilde{\gamma}$ is always correspondingly adjusted so that $\tilde{\gamma}=\Omega/2$. Both the maximum achievable entanglement and the size of the parameter region with nonzero entanglement rapidly decreases with increasing bandwidth of the frequency noise spectrum. Again, entanglement is more robust against phase noise in the resolved sideband limit $\kappa/\omega_{\rmm} < 1$.

\subsection{Approximate analytical expressions for $E_N$}

The exact expression of the logarithmic negativity stemming from the solution of Eq.~(\ref{lyapunov}) is cumbersome, but it is nonetheless possible to explain the above results by means of an approximate treatment which satisfactorily describes the effect of phase noise on $E_N$. In fact, by following the approach of Ref.~\cite{Genes2008b}, based on the solution of Eqs.~(\ref{linear}) in the frequency domain, for each matrix element of the $4 \times 4$ stationary CM one can write
\begin{equation}
V_{ij} = \int_{-\infty}^{\infty} \frac{{\rm d }\omega}{2 \pi}\left[V_{ij}(\omega)+ \delta V_{ij}(\omega)\right],  \\
\end{equation}
where $V_{ij}(\omega)$ refers to the situation with no laser noise and $\delta V_{ij}(\omega)$ is the correction due to the presence of laser phase noise. This latter correction is explicitly given by
\begin{equation}\label{phasecorrection}
   \delta V_{ij}(\omega)=\frac{2 |\alpha_{\rm s}|^2  |\chi_{\mathrm{eff}}(\omega)|^2 {\mathcal S}_{\dot{\phi}}(\omega)c_i(\omega)c_j(\omega)^*}{[\kappa^2+(\omega-\Delta)^2][\kappa^2+(\omega+\Delta)^2]} ,
\end{equation}
where
\begin{equation}
\chi_{\mathrm{eff}}(\omega)= \left[ \omega_{\rmm}^2-\omega^2- \rmi \gamma_{\rmm} \omega -\frac{G^2 \Delta \omega_{\rmm}}{(\kappa-\rmi \omega)^2+\Delta^2} \right]^{-1},
\label{susceptibility}
\end{equation}
is the effective susceptibility of the mechanical oscillator modified by radiation pressure \cite{Genes2008}, and $c(\omega)$ is
the vector
\begin{eqnarray*}
&& c(\omega) =\left[G \Delta \omega_{\rm m}, -i \omega G \Delta, \Delta (\omega_{\rm m}^2-\omega^2-i\gamma_m \omega),\right.\\
&& \left.(\kappa-i\omega) (\omega_{\rm m}^2-\omega^2-i\gamma_m \omega)\right]^{\mathsf{T}}.
\end{eqnarray*}
As discussed in Ref.~\cite{Genes2008} (see also Ref.~\cite{Dantan2008}), $|\chi_{\mathrm{eff}}(\omega)|^2$ has a Lorenzian-like form, peaked at an effective frequency $\omega_{\rmm}^{\rm eff}$, and with width given by an effective damping rate $\gamma_{\rmm}^{\rm eff}$, whose approximate expressions are given by \cite{Genes2008}
\begin{eqnarray}
&&\omega _{\rm m}^{\rm eff}=\left[ \omega _{\rm m}^{2}-\frac{G^{2}\Delta \omega
_{\rm m}(\kappa ^{2}-\omega_{\rm m} ^{2}+\Delta ^{2})}{\left[ \kappa ^{2}+(\omega_{\rm m}
-\Delta )^{2}\right] \left[ \kappa ^{2}+(\omega_{\rm m} +\Delta )^{2}\right] }\right]
^{\frac{1}{2}},  \label{omegeff} \\
&&\gamma _{\rm m}^{\rm eff}=\gamma _{\rm m}+\frac{2G^{2}\Delta \omega_{\rm m}
\kappa }{\left[ \kappa ^{2}+(\omega_{\rm m} -\Delta )^{2}\right] \left[
\kappa ^{2}+(\omega_{\rm m} +\Delta )^{2}\right] } \label{dampeff}.
\end{eqnarray}
The modification of the mechanical frequency due to radiation pressure shown by Eq.~(\ref{omegeff}) is the so-called ``optical spring effect'' which may lead to significant frequency shifts in the case of low-frequency oscillators \cite{Corbitt2007}. In the case of higher resonance frequencies (around $1$ MHz) and red-detuned laser, as we are assuming here, the modification can be appreciable only very close to the bistability threshold.

In a wide parameter region, and especially in the resolved sideband regime $\kappa < \omega_m$, the integrand in Eq.~(\ref{phasecorrection}) is dominated by the peak of $|\chi_{\mathrm{eff}}(\omega)|^2$ and therefore one can evaluate the corrections $\delta V_{ij}$ by approximating the laser phase noise spectrum with a constant given by its value at the peak, ${\mathcal S}_{\dot{\phi}}(\omega_{\rmm}^{\rm eff})$. This means that in the resolved sideband regime, the effect of laser phase noise on the stationary state of the system is completely described by a unique number, the phase noise spectrum at the effective mechanical frequency ${\mathcal S}_{\dot{\phi}}(\omega_{\rmm}^{\rm eff})$. Therefore the effects of laser phase noise can be minimized simply by suppressing the noise spectrum at $\omega_{\rmm}^{\rm eff}$, and this fact explains why the explicit form of the phase noise spectrum is relevant.
We have verified that the approximated CM obtained by replacing ${\mathcal S}_{\dot{\phi}}(\omega)$ with ${\mathcal S}_{\dot{\phi}}(\omega_{\rmm}^{\rm eff})$ in the phase noise terms of Eq.~(\ref{phasecorrection}) satisfactorily reproduces the behavior of $E_N$ shown in Figs.~2-5, especially in the resolved sideband regime.

Such an approximation for the laser phase noise contribution to the CM can also be used to derive an approximate analytical expression for $E_N$ in the parameter region very close to the bistability threshold. This regime is relevant for optomechanical entanglement because, as first pointed out in Refs.~\cite{Genes2008b,Genes2009} and recently discussed in detail in Ref.~\cite{Ghobadi2011}, $E_N$ reaches its maximum value at the bistability threshold in the ideal case with no laser noise. As shown in Figs.~2-5, this is no more true in the presence of phase noise, when, on the contrary, $E_N$ always drops to zero close to threshold. This is well explained by the approximate expression of the symplectic eigenvalue $\eta^-$ which is obtained by approximating ${\mathcal S}_{\dot{\phi}}(\omega)$ with ${\mathcal S}_{\dot{\phi}}(\omega_{\rmm}^{\rm eff})$ in Eq.~(\ref{phasecorrection}), taking for $G$ the threshold value $ G \simeq \sqrt{(\kappa^2+\Delta^2)\omega_{\rm m}/\Delta}$, and neglecting thermal noise terms. One gets
\begin{equation}\label{approsymple}
\eta^{-}\simeq \frac{1}{\sqrt{2}}\sqrt{\frac{a+b \mathcal{S}_{\dot{\phi}}(\omega_{\rmef}) +c \mathcal{S}_{\dot{\phi}}(\omega_{\rmef})^2 +d \mathcal{S}_{\dot{\phi}}(\omega_{\rmef})^3}{f+g \mathcal{S}_{\dot{\phi}}(\omega_{\rmef})}},
\end{equation}
where
\begin{eqnarray}
a&=&\kappa^3 (\kappa^2 +\Delta^2)\left[4 \Delta^4 +4 \Delta^2 \left(\kappa^2 +\omega_{\rmm}^2 \right)+\omega_{\rmm}^4\right], \\
b&=&2|\alpha_s|^2 \Delta^2 \kappa^2 \left[4 (\Delta^2 + \kappa^2) (2 \Delta^2 + \kappa^2) \right.\nonumber \\
&& \left.+6 (\Delta^2 +\kappa^2)\omega_{\rmm}^2 + \omega_{\rmm}^4\right], \\
c&=&4 |\alpha_s|^4 \Delta^4 \kappa [5 (\Delta^2 + \kappa^2) + 2 \omega_{\rmm}^2], \\
d&=&8 |\alpha_s|^6 \Delta^6, \\
f&=&8 \kappa^3 \Delta ^2 (\kappa^2 +\Delta^2)\left(\Delta ^2+\kappa ^2+5 \omega _m^2\right),\\
g&=&16 |\alpha_s|^2 \kappa^2 \Delta^4 \left(\Delta^2+ \kappa^2 +\omega_{\rmm}^2\right).
\end{eqnarray}
If laser phase noise is negligible, $\mathcal{S}_{\dot{\phi}}(\omega_{\rmef})=0$, one gets (see also Ref.~\cite{Ghobadi2011})
\begin{equation}\label{en-nophasenoise}
\eta^{-} \simeq \sqrt{\frac{a}{2f}} =\sqrt{\frac{4 \Delta^4 +4 \Delta^2 \left(\kappa^2 +\omega_{\rmm}^2 \right)+\omega_{\rmm}^4}{16 \Delta^2 \left(\Delta^2 +\kappa^2 +5 \omega_{\rmm}^2 \right)}}.
\end{equation}
This value is minimum at $$\Delta/\omega_{\rm m}=\frac{1}{4}\sqrt{1+\sqrt{\left(\frac{4 \kappa}{\omega_{\rm m}}\right)^2+81}},$$ at which $E_N$ achieves its maximum value $$E_N = -\ln \left[\frac{1}{5}\sqrt{9+\frac{128 \kappa ^2}{8 \kappa^2+45\omega_{\rm m}^2}}\right],$$ which can become at most $\ln[5/3] \simeq 0.51$ in the resolved sideband limit $\kappa \ll \omega_{\rm m}$ \cite{Ghobadi2011}.

In the presence of phase noise instead, the terms proportional to $\mathcal{S}_{\dot{\phi}}(\omega_{\rmef})$ and its powers in Eq.~(\ref{approsymple}) are always predominant because $|\alpha_s|^2$ is typically large, and therefore $\eta^-$ may easily become larger than $1/2$, which means no entanglement, even for not too large values of the laser linewidth $\Gamma_l$. This shows why entanglement vanishes close to the bistability threshold as soon as phase noise is present, and also why the maximum $E_N$ is attained at smaller input power for increasing phase noise.

\subsection{Cooling}

The effect of laser phase noise on ground state cooling of the mechanical resonator has been already discussed by various papers \cite{Diosi2008,Yin2009,Rabl2009,Phelps2011}. Ref.~\cite{Rabl2009} in particular provided an accurate estimation of the effect of laser noise and showed that the relevant quantity is just the frequency noise spectrum at the mechanical resonance, $\mathcal{S}_{\dot{\phi}}(\omega_{\rmm})$: laser phase noise does not pose serious limitations to cooling provided that $\mathcal{S}_{\dot{\phi}}(\omega_{\rmm})$ is not too large. The analysis of the effect of phase noise on the stationary CM in the preceding subsection fully confirms such a prediction, i.e., we recover the results of Ref.~\cite{Rabl2009} on ground state cooling, even though by means of a different treatment, based on QLE instead of the master equation, and working in the frame rotating at the fluctuating frequency.

In fact, the stationary mean energy of the mechanical oscillator is given by
\begin{equation}
U =\frac{\hbar \omega_{\rmm}}{2} [\langle \delta q^2 \rangle + \langle \delta p^2 \rangle] \equiv \hbar \omega_{\rmm}(n_{\mathrm{eff}}+\frac{1}{2}),
\label{mecenergy}
\end{equation}
i.e., it is a linear combination of the stationary CM matrix elements $V_{11}$ and $V_{22}$. Therefore one can again exploit the fact that $|\chi_{\mathrm{eff}}(\omega)|^2$ is strongly peaked, and approximate the frequency integrals for $\langle \delta q^2 \rangle$ and $\langle \delta p^2 \rangle$ by replacing ${\mathcal S}_{\dot{\phi}}(\omega)$ with ${\mathcal S}_{\dot{\phi}}(\omega_{\rmm}^{\rm eff})$.
Assuming also the weak coupling regime $\kappa \gg \gamma_{\rmm}, G$ and $\omega_{\rmm} \gg \overline{n} \gamma_{\rmm}, G$ which is the relevant one for ground state cooling, one obtains a result analogous to that of Ref.~\cite{Rabl2009}, which generalizes the expression of Refs.~\cite{Marquardt2007,Wilson-Rae2007,Genes2008} by including the effect of laser phase noise,
\begin{equation}
n_{\mathrm{eff}}=\frac{1}{\gamma_{\rmm}+\Gamma_{\mathrm{op}}} \left[ n \gamma_{\rmm} + A_{+}
+ \frac{|\alpha_s|^2 \Delta \Gamma_{\mathrm{op}}}{2 \kappa \omega_{\rmm}} \mathcal{S}_{\dot{\phi}}(\omega_{\rmm}^{\rm eff}) \right].
\label{neff}
\end{equation}
Here we denote with $\Gamma_{\mathrm{op}} = A_{-}-A_{+}$ the net laser cooling rate, and
\begin{equation}
A_{\pm} = \frac{\kappa G^2}{2[\kappa^2 +(\Delta \pm \omega_{\rmm})^2]}
\label{apm}
\end{equation}
denotes the two scattering rates of laser photons into the Stokes sideband ($A_+$) or the anti-Stokes sideband ($A_-$) of the laser. Recall that, as pointed out in Ref.~\cite{Genes2008}, $\gamma_{\rmm}+\Gamma_{\mathrm{op}}= \gamma _{\rm m}^{\rm eff}$, i.e., it coincides with the effective mechanical damping of Eq.~(\ref{dampeff}).

Eq.~(\ref{neff}) reproduces Eq.~(23) of Ref.~\cite{Rabl2009} if we recall that in typical situations $\Gamma_{\mathrm{op}} \gg \gamma_{\rmm}$ and by restricting to the optimal condition for ground state cooling, i.e., the resolved sideband limit $G,\kappa \ll \omega_{\rmm}$ and $\Delta \approx \omega_{\rmm}$, where the optical spring effect is negligible and $\omega_{\rmm}^{\rm eff} \simeq \omega_{\rmm}$. We have compared the prediction of Eq.~(\ref{neff}) with the exact solution of the stationary state of the system, and we have verified that it works very well in the relevant regime where ground state cooling is achievable. The behavior of $n_{\rm eff}$ for the same set of parameter values considered above is studied in Figs.~6-9.
\begin{figure*}
\centering
\subfigure{
\includegraphics[width=2.34in]{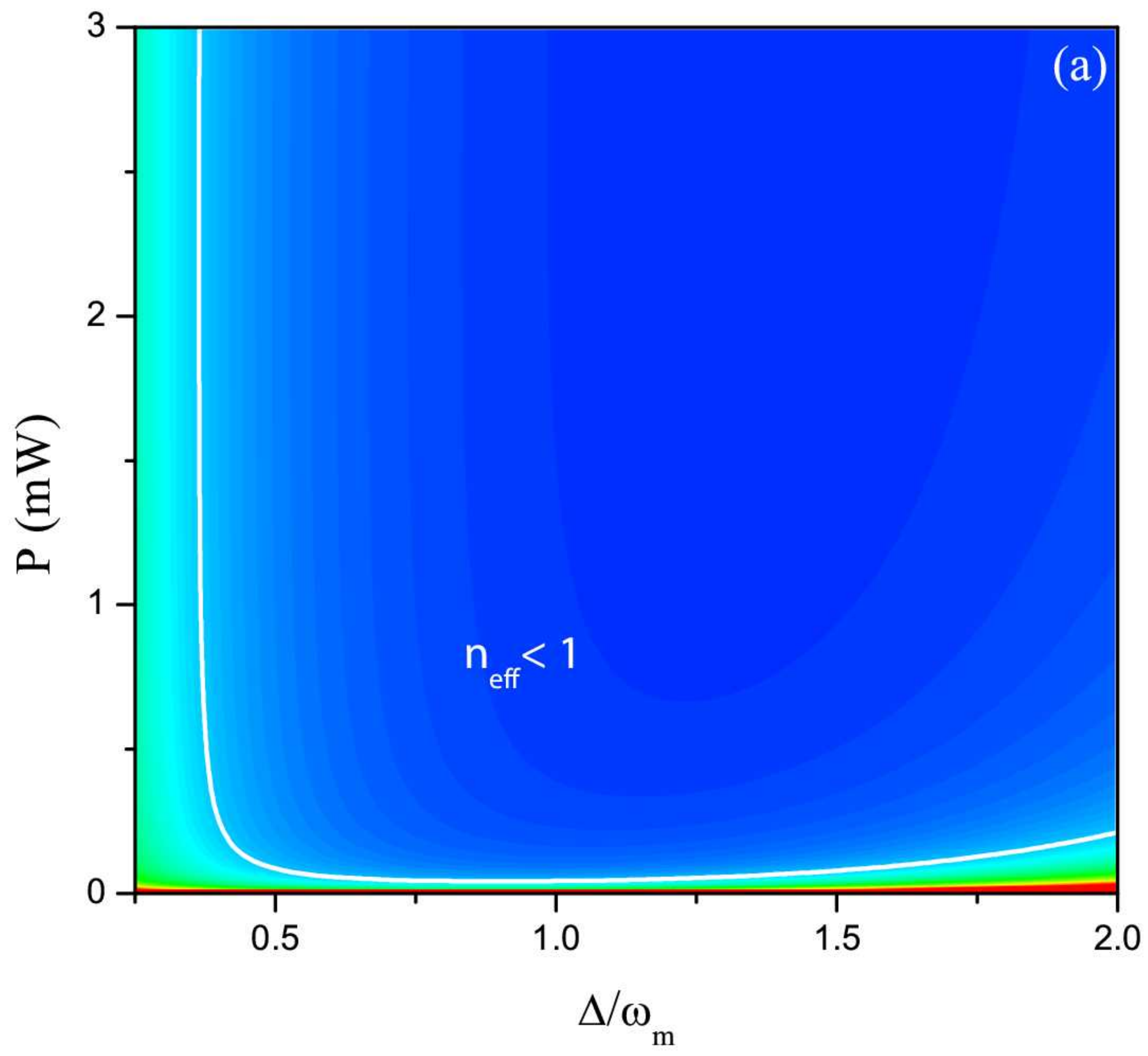}
\label{fig:nefdetlin_a}
}
\subfigure{
\includegraphics[width=2.07in]{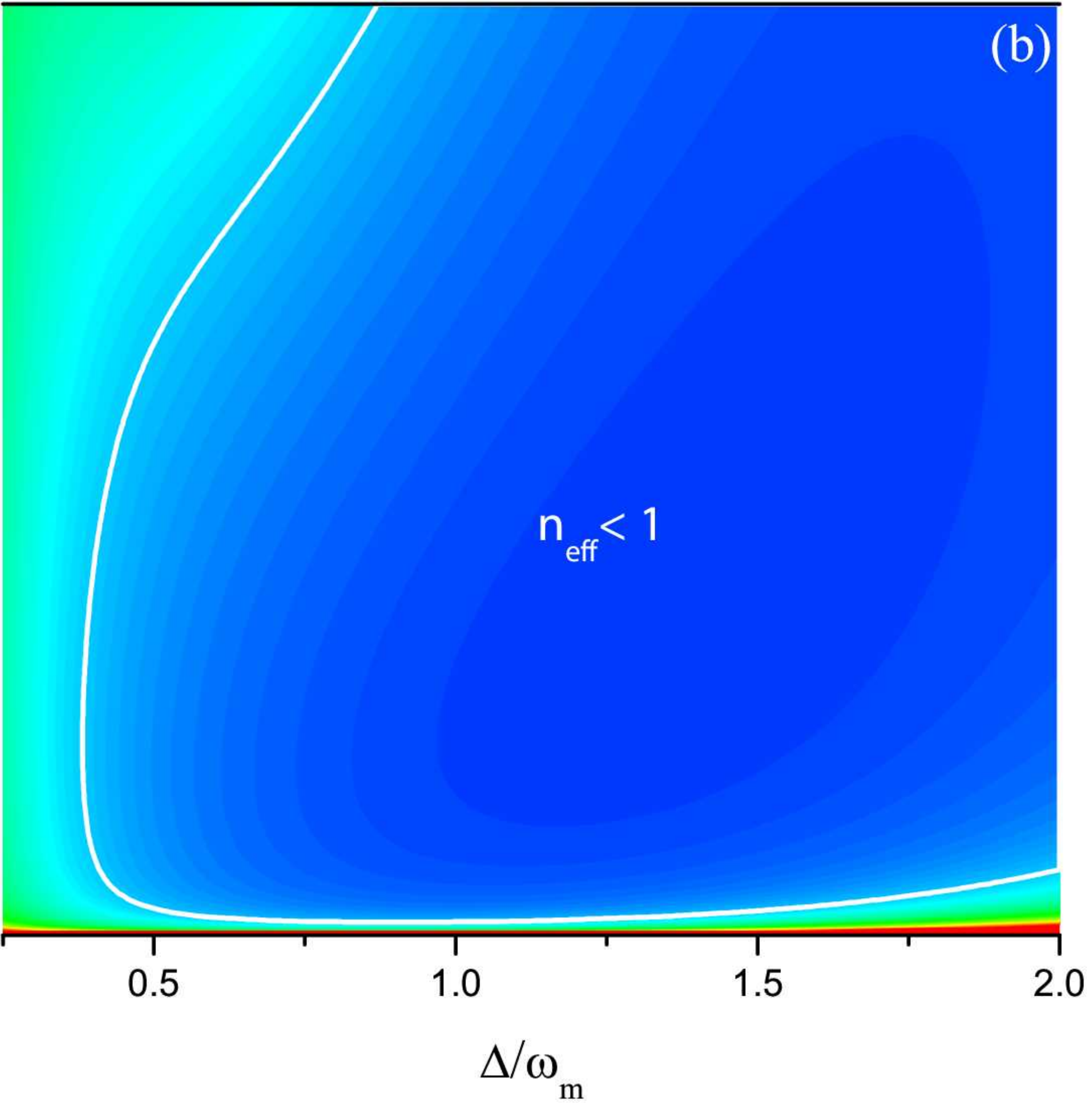}
\label{fig:nefdetlin_b}
}
\subfigure{
\includegraphics[width=2.31in]{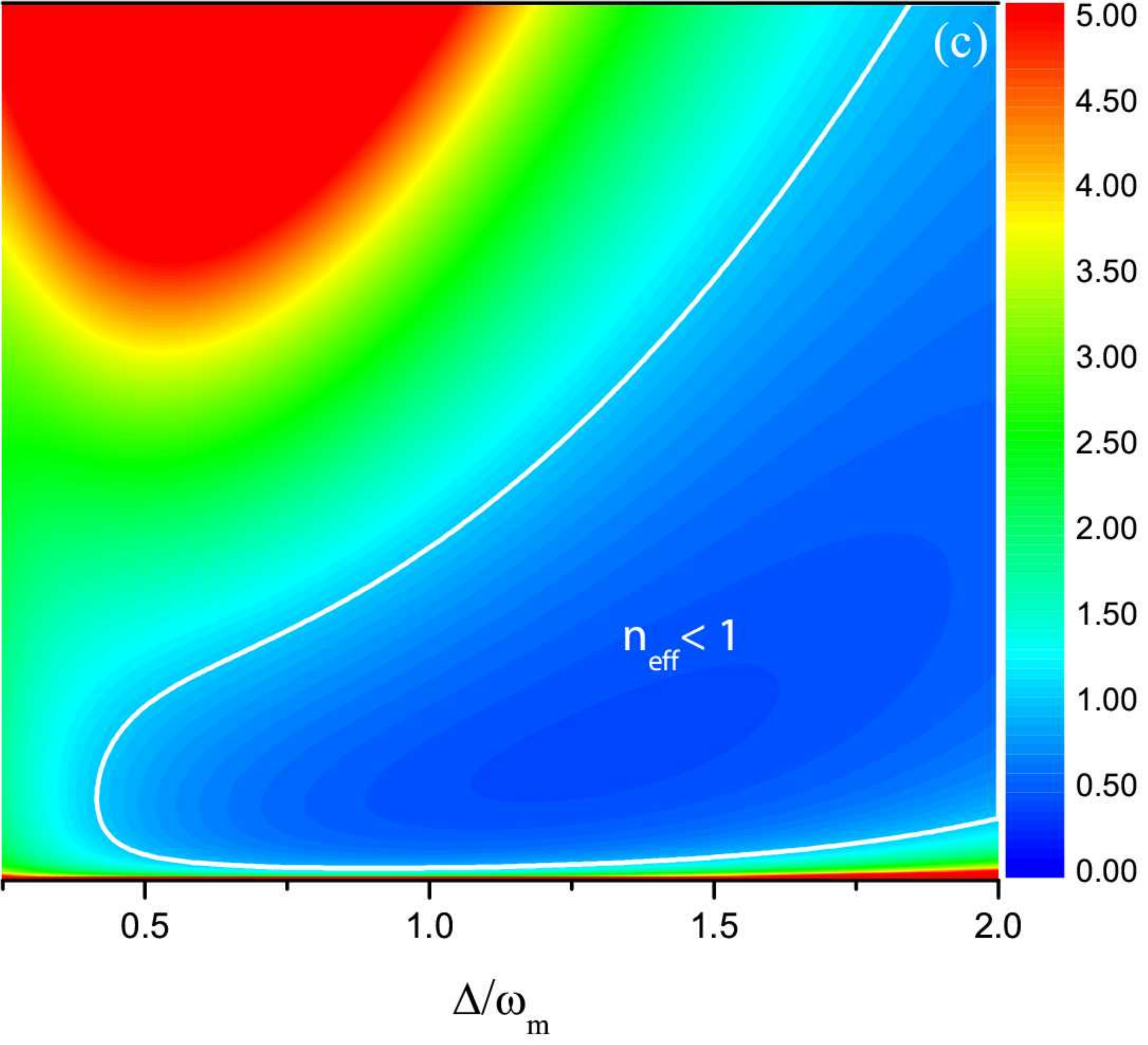}
\label{fig:nefdetlin_c}
}
\caption{(Color online) Contour plot of $n_{{\rm eff}}$ versus the input power $P$ and normalized cavity detuning $\Delta/\omega_{\rmm}$ for $\Gamma_l=0$ (a), $\Gamma_l/2\pi=0.1$ KHz (b), and $\Gamma_l/2\pi=1$ KHz (c). We have also fixed $\kappa/\omega_{\rmm}=1$ and $\Omega/2\pi=50$ KHz and the other parameter values are fixed in the text.}
\label{fig:nefdetlin}
\end{figure*}
\begin{figure*}
\centering
\subfigure{
\includegraphics[width=2.33in]{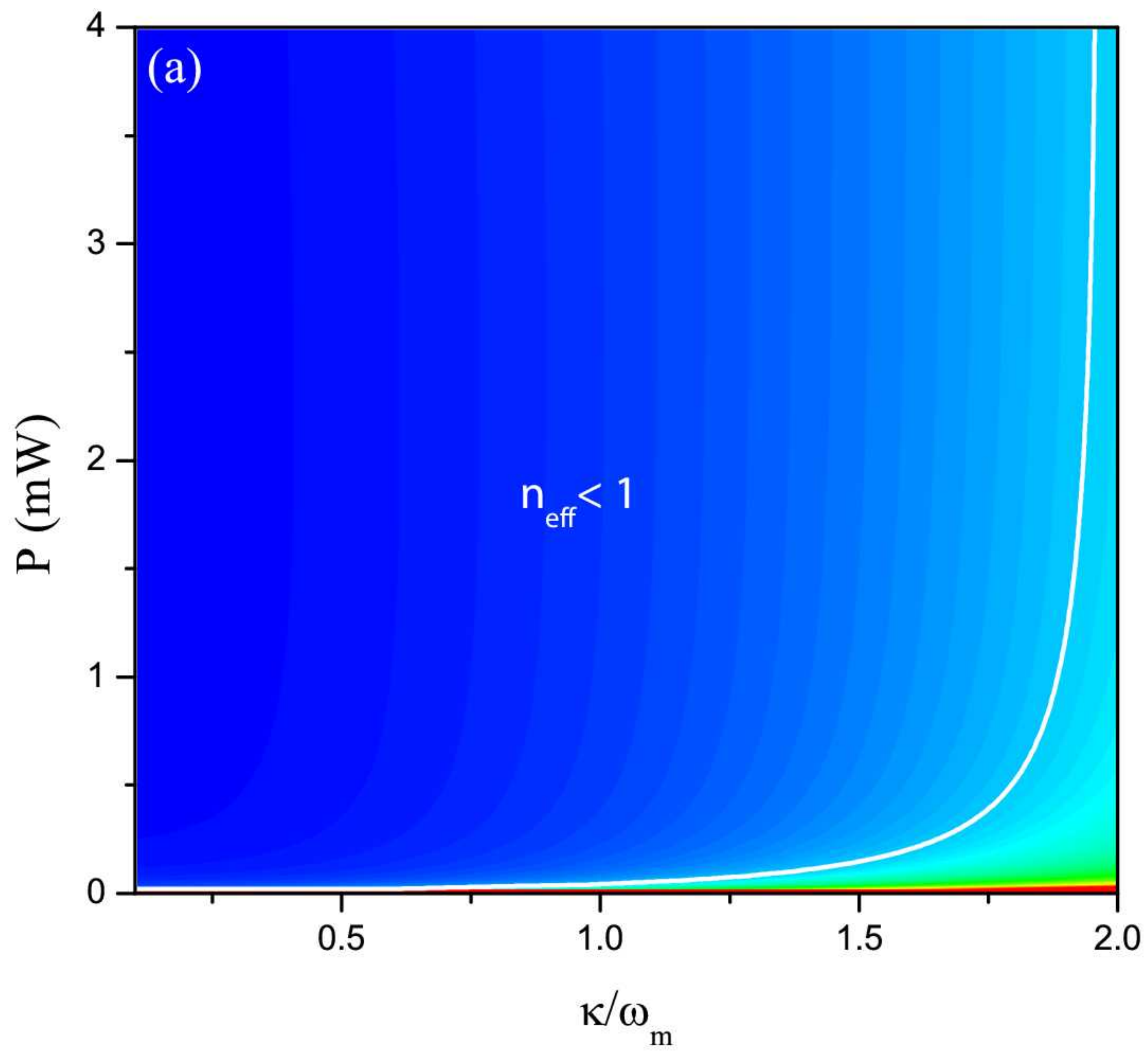}
\label{fig:neffinlin_a}
}
\subfigure{
\includegraphics[width=2.06in]{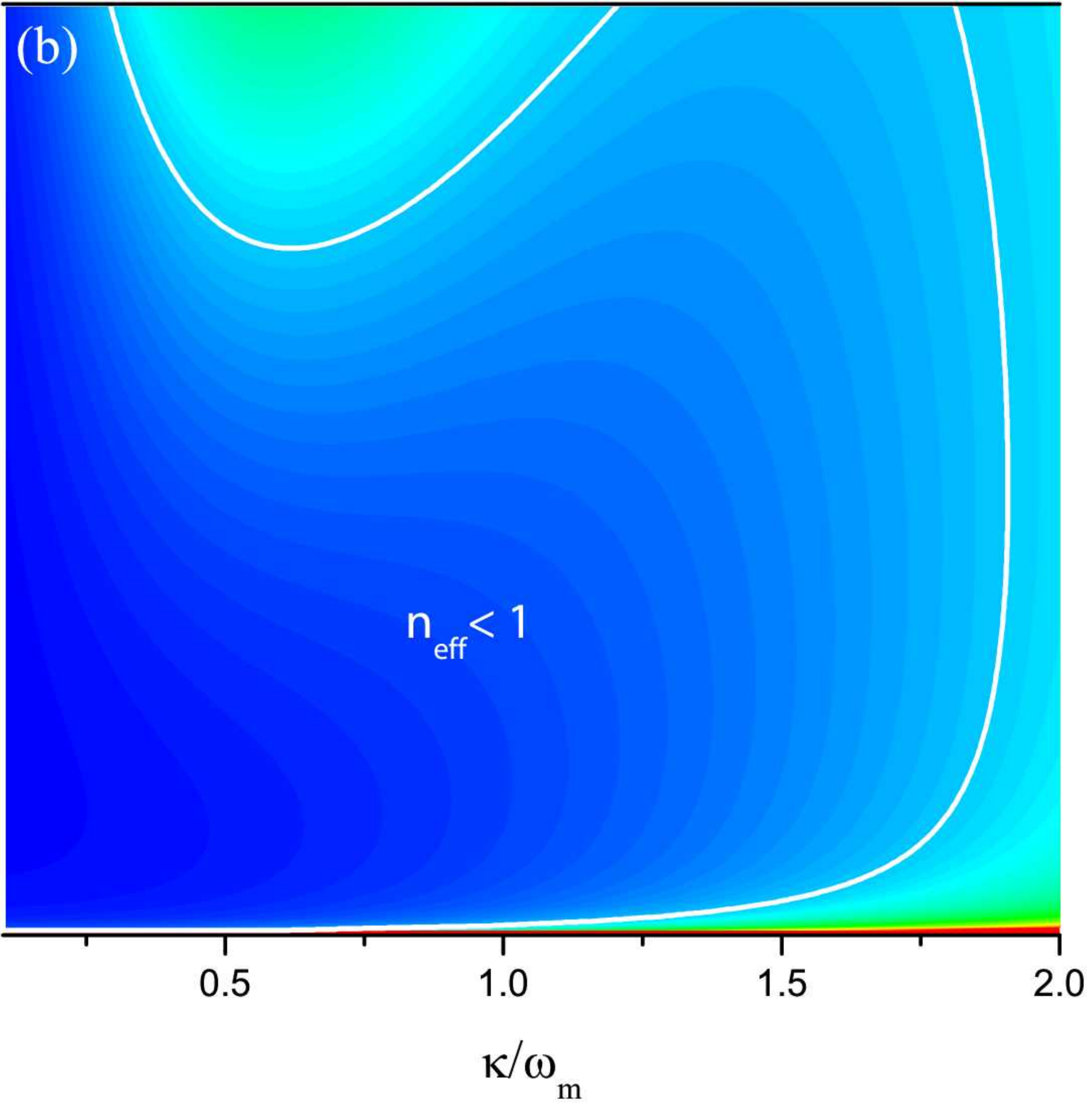}
\label{fig:neffinlin_b}
}
\subfigure{
\includegraphics[width=2.31in]{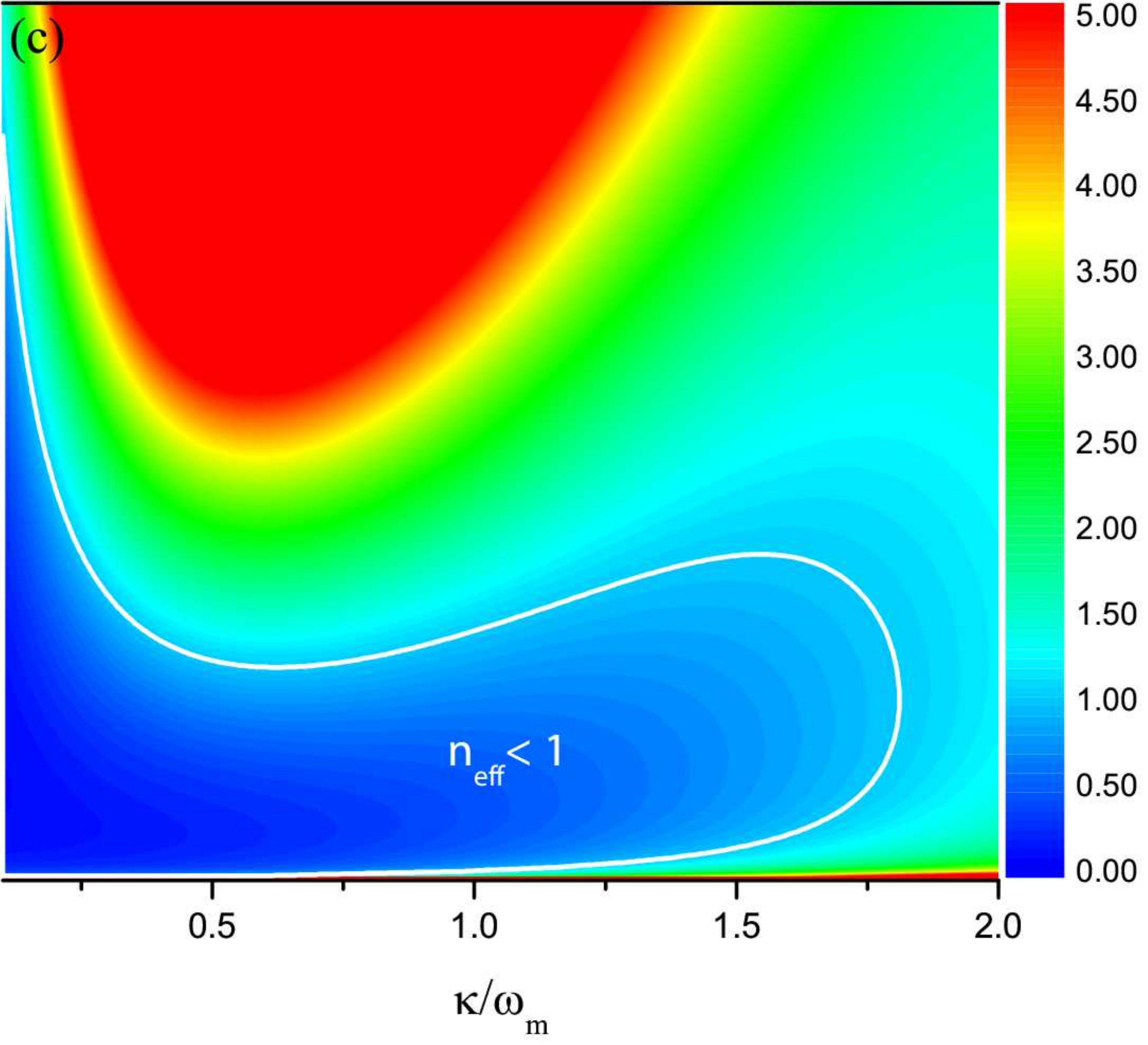}
\label{fig:neffinlin_c}
}
\caption{(Color online) Contour plot of $n_{{\rm eff}}$ versus the input power $P$ and normalized cavity bandwidth $\kappa/\omega_{\rmm}$ for $\Gamma_l=0$ (a), $\Gamma_l/2\pi=0.1$ KHz (b), and $\Gamma_l/2\pi=1$ KHz (c). We have also fixed $\Delta/\omega_{\rmm}=1$ and $\Omega/2\pi=50$ KHz and the other parameter values are fixed in the text.}
\label{fig:neffinlin}
\end{figure*}

In Fig.~6 we show $n_{{\rm eff}}$ versus the input power $P$ and normalized cavity detuning $\Delta/\omega_{\rmm}$ for $\Gamma_l=0$ (a), $\Gamma_l/2\pi=0.1$ KHz (b), and $\Gamma_l/2\pi=1$ KHz (c). The cavity bandwidth has been fixed at $\kappa/\omega_{\rmm}=1$. Fig.~7 shows instead $n_{{\rm eff}}$ versus the input power $P$ and normalized cavity decay rate $\kappa/\omega_{\rmm}$ again for $\Gamma_l=0$ (a), $\Gamma_l/2\pi=0.1$ KHz (b), and $\Gamma_l/2\pi=1$ KHz (c). The cavity detuning is now fixed at $\Delta/\omega_{\rmm}=1$, and in both figures we have fixed $\Omega/2\pi=50$ KHz. The effective occupancy is less affected than entanglement by laser phase noise: in fact for increasing $\Gamma_l$ the parameter region where $n_{{\rm eff}} < 1$ becomes narrower but one can still reach ground state cooling for a realistic set of parameters, provided that the input power is not too large. Similarly to what happens for entanglement, ground state cooling is achievable only in the resolved sideband limit for increasing laser noise strength $\Gamma_l$ (see Fig.~\ref{fig:neffinlin_c}).
\begin{figure*}
\centering
\subfigure{
\includegraphics[width=2.34in]{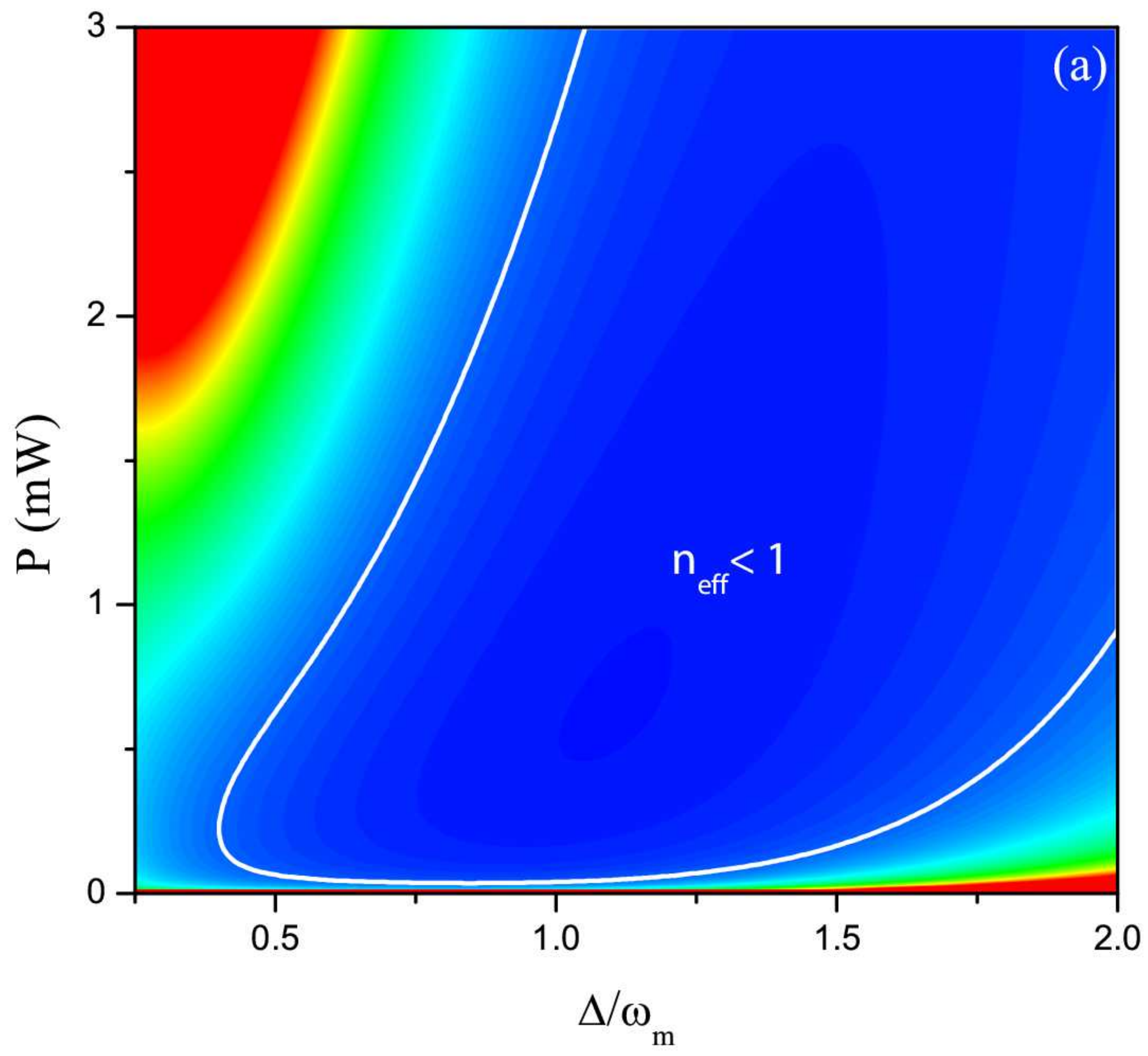}
\label{fig:nefdetcut_a}
}
\subfigure{
\includegraphics[width=2.07in]{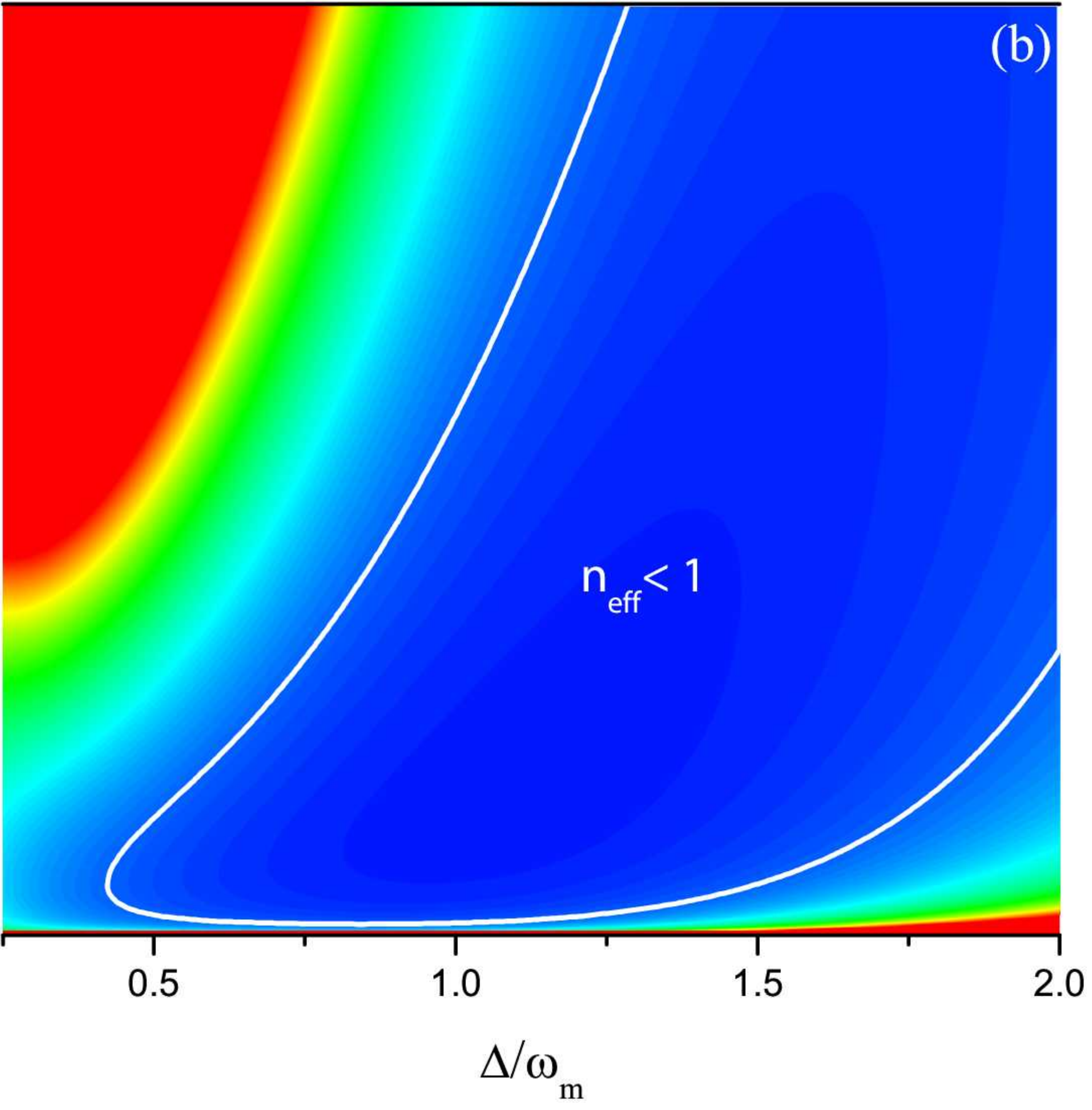}
\label{fig:nefdetcut_b}
}
\subfigure{
\includegraphics[width=2.31in]{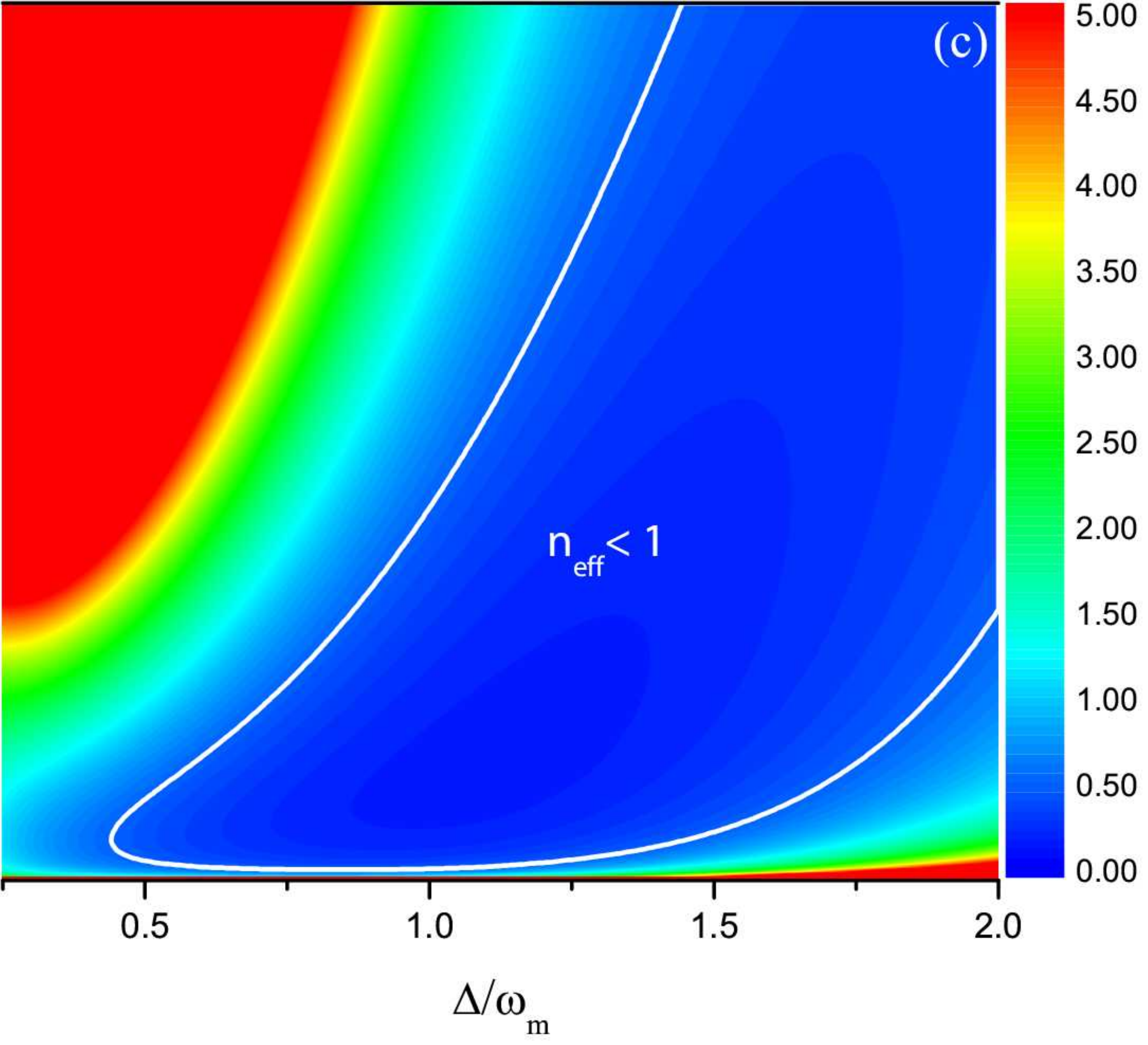}
\label{fig:nefdetcut_c}
}
\caption{(Color online) Contour plot of $n_{{\rm eff}}$ versus the input power $P$ and normalized cavity detuning $\Delta/\omega_{\rmm}$ at fixed laser linewidth $\Gamma_l/2\pi=0.1$ KHz, fixed bandwidth $\kappa/\omega_{\rmm}=1$, and for different values of the center of the noise spectrum, $\Omega /2\pi =30$ KHz (a), $\Omega/2\pi =80$ KHz (b), and $\Omega/2\pi =140$ KHz (c) (the bandwidth parameter $\tilde{\gamma}$ is always adjusted so that $\tilde{\gamma}=\Omega/2$).}
\label{fig:nefdetcut}
\end{figure*}
\begin{figure*}
\centering
\subfigure{
\includegraphics[width=2.33in]{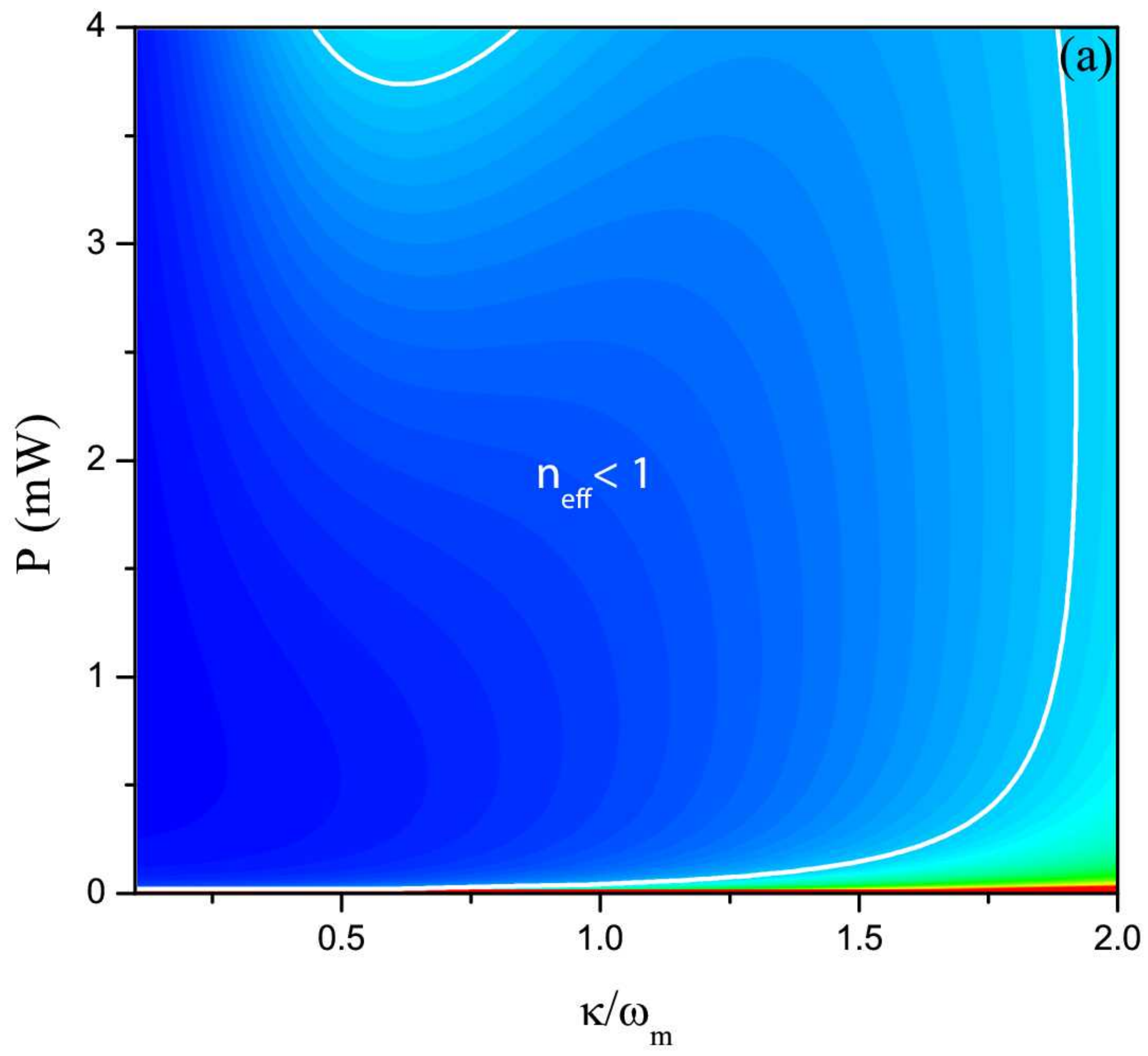}
\label{fig:neffincut_a}
}
\subfigure{
\includegraphics[width=2.06in]{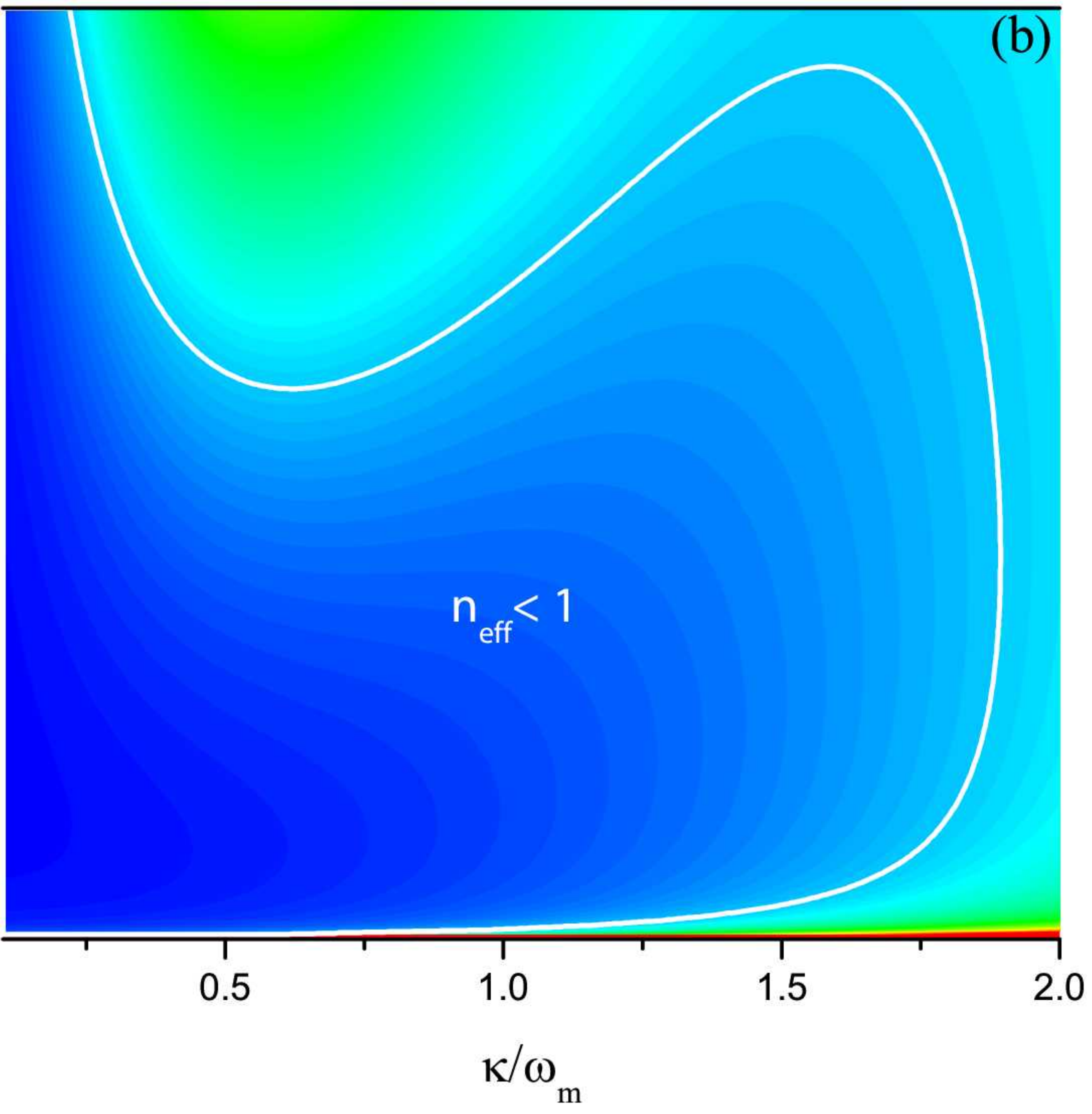}
\label{fig:neffincut_b}
}
\subfigure{
\includegraphics[width=2.31in]{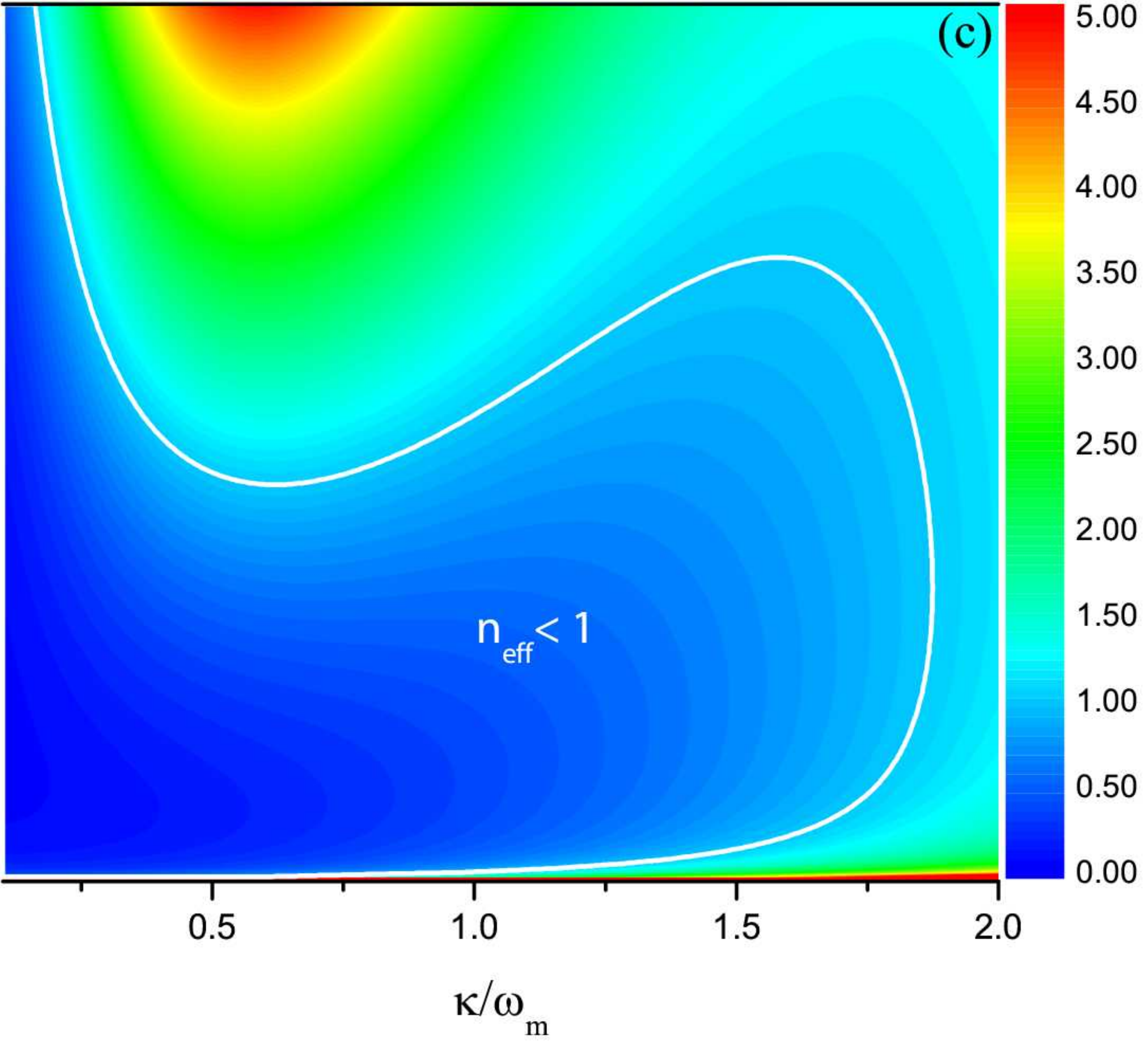}
\label{fig:neffincut_c}
}
\caption{(Color online) Contour plot of $n_{{\rm eff}}$ versus the input power $P$ and normalized cavity bandwidth $\Delta/\omega_{\rmm}$ at fixed laser linewidth $\Gamma_l/2\pi=0.1$ KHz, fixed detuning $\kappa/\omega_{\rmm}=1$, and for different values of the center of the noise spectrum, $\Omega /2\pi =30$ KHz (a), $\Omega/2\pi =80$ KHz (b), and $\Omega/2\pi =140$ KHz (c) (the bandwidth parameter $\tilde{\gamma}$ is always adjusted so that $\tilde{\gamma}=\Omega/2$).}
\label{fig:neffincut}
\end{figure*}

In Figs.~8 and 9 instead we study the dependence of the phonon occupancy upon the spectral properties of laser phase noise. Fig.~8 shows $n_{{\rm eff}}$ versus the input power $P$ and normalized cavity detuning $\Delta/\omega_{\rmm}$ at fixed laser linewidth $\Gamma_l/2\pi=0.1$ KHz, fixed bandwidth $\kappa/\omega_{\rmm}=1$, and for different values of $\Omega$, $\Omega/2\pi =30$ KHz (a), $\Omega/2\pi =80$ KHz (b), and $\Omega/2\pi =140$ KHz (c) (the bandwidth parameter $\tilde{\gamma}$ is always adjusted so that $\tilde{\gamma}=\Omega/2$). In Fig.~9 instead we plot $n_{{\rm eff}}$ versus the input power $P$ and normalized cavity decay rate $\kappa/\omega_{\rmm}$, at fixed $\Delta/\omega_{\rmm}=1$ and for the same set of values for $\Gamma_l$ and $\Omega$. As predicted by Eq.~(\ref{neff}), it is just the noise spectrum at the effective mechanical resonance frequency which mainly affects cooling. When $\Omega$ (and consequently $\tilde{\gamma}$) is increased, the laser noise spectrum broadens and its value at $\omega_{\rmm}^{\rm eff}$ increases as well. As a consequence, the parameter region where $n_{\rm eff} < 1$ becomes narrower and narrower. In particular for a broader spectrum ground state cooling is better achieved at not too large values of input power and again in the resolved sideband limit $\kappa/\omega_{\rmm} < 1$.

%
%
\section{Conclusion}

We have studied the effects of the fluctuations of both the amplitude and the phase of the laser driving a cavity optomechanical system. We have analyzed the dynamics by adopting a quantum Langevin treatment, in which the phase noise dynamics has been included by means of additional auxiliary variables. We have linearized the dynamics around the classical stationary state of the system and analyzed the dependence of the log-negativity of the stationary state and of the mechanical occupancy upon the various system parameters. We have also derived approximate, but compact analytical expressions showing the effect of laser phase noise on these quantities. We have seen that, even though laser noise may have an appreciable affect on the quantum properties of the steady state, both cooling to the mechanical ground state and stationary optomechanical entanglement are still possible if state-of-the-art stable lasers are employed.

\section{Acknowledgments}

This work has been supported by the European Commission (FP-7 FET-Open project MINOS), and by INFN (SQUALO project).

%

\bibliography{optomechanics}

\end{document}